%% file: main.tex
\documentclass[10pt,conference]{IEEEtran}

\AtBeginDocument{%
  \providecommand\BibTeX{{%
    \normalfont B\kern-0.5em{\scshape i\kern-0.25em b}\kern-0.8em\TeX}}}

\input{macro}

\begin{document}



\title{Beyond Static GUI Agent: Evolving LLM-based GUI Testing via Dynamic Memory}


\author{
Mengzhuo Chen$^{1,2,3,*}$, Zhe Liu$^{1,2,3,*}$, Chunyang Chen$^4$, Junjie Wang$^{1,2,3,\dagger}$, \\
Yangguang Xue$^{1,2,3}$, Boyu Wu$^{1,2}$, Yuekai Huang$^{1,2,3}$, Libin Wu$^{1,2,3}$, Qing Wang$^{1,2,3,\dagger}$ \\
\IEEEauthorblockA{$^1$Institute of Software Chinese Academy of Sciences, Beijing, China; \\ $^2$University of Chinese Academy of Sciences, Beijing, China; \\ $^3$ State Key Laboratory of Complex System Modeling and Simulation Technology ISCAS, Beijing, China; \\ $^4$Technical University of Munich, Munich, Germany;\\
$^*$Both authors contributed equally to this research. $\dagger$Corresponding author; \\ \{chenmengzhuo23, liuzhe181\}@mails.ucas.edu.cn, junjie@iscas.ac.cn, wq@iscas.ac.cn
}}


\maketitle

\begin{abstract}
The development of Large Language Models (LLMs) enables LLM-based GUI testing to interact with graphical user interfaces by understanding GUI screenshots and generating actions, which are widely applied in industry and academia. However, current approaches test each app in isolation, lacking mechanisms for experience accumulation and reuse. 
This limitation often causes GUI testing approaches to miss deeper exploration and fail to trigger bug-prone functionalities.
To address this, we propose {\tool}, a three-layer memory mechanism that augments LLM-based GUI testing with the ability to evolve through repeated interaction. {\tool} designs episodic memory to capture functional-level testing traces, reflective memory to summarize issue patterns and redundant behaviors, and strategic memory to synthesize cross-app exploration strategies. These memory layers are dynamically retrieved and injected into LLM prompts at runtime, enabling the agent to reuse successful behaviors, avoid ineffective actions, and prioritize bug-prone paths.
We implement {\tool} as a lightweight plugin, which can be integrated into existing LLM-based GUI testing approaches.
We evaluate {\tool} on real-world apps from 15 diverse app categories. Results show that {\tool} enhances GUI testing performance across five baselines, with activity and code coverage increasing by 79\% - 96\% and 81\% - 97\%, and bug detection improving by 57\% - 198\%. Ablation studies confirm the contributions of each memory layer. Furthermore, {\tool} detects 49 new bugs in 200 popular apps, with 35 confirmed fixes and 14 acknowledged by developers, showing its practical value in memory-driven GUI testing.
\end{abstract}


\input{sec/introduction}

\input{sec/motivation}
\input{sec/approach}

\input{sec/experiment}

\input{sec/result}

\input{sec/discussion}

\input{sec/related}
\input{sec/conclusion}

\section*{Acknowledgment}
This work was supported by the National Natural Science Foundation
of China Grant No. 62232016, 62402483, 62072442, 62402484,
Major Program of ISCAS Grant No. ISCAS-ZD-202401 and ISCAS-ZD-202302, Innovation
Team 2024 ISCAS (No. 2024-66), Basic Research Program of ISCAS Grant No. ISCAS-JCZD-202304.






\bibliographystyle{IEEEtran}


\bibliography{reference}

\end{document}

%% file: macro.tex
\usepackage[linesnumbered,ruled,vlined]{algorithm2e}
\usepackage{flushend}

\usepackage{amsmath}

\usepackage[table]{xcolor}
\usepackage{soul}
\usepackage{float}
\usepackage{threeparttable}
\usepackage{booktabs}
\usepackage{multirow}
\usepackage{epstopdf}
\usepackage{hyperref}
\usepackage{listings}
\usepackage{fancybox}
\usepackage{graphicx}
\usepackage{subcaption}
\usepackage{paralist}
\usepackage{ragged2e}
\usepackage{enumitem}
\usepackage{balance}

\usepackage{marvosym}
\usepackage{multirow}
\usepackage{epstopdf}
\usepackage{hyperref}
\usepackage{listings}
\usepackage{fancybox}
\usepackage{graphicx}
\usepackage{subcaption}
\usepackage{subfloat}
\usepackage{cleveref}
\usepackage{paralist}
\usepackage{ragged2e}
\usepackage{enumitem}
\usepackage{pythonhighlight}
\usepackage{algpseudocode}

\usepackage{cite}

\usepackage{minted}
\usepackage{caption}
\usepackage{xparse}

\newminted{c}{
    linenos, 
    frame=lines, 
    tabsize=4, 
    autogobble,
}

\newminted{java}{
    linenos, 
    frame=lines, 
    tabsize=4, 
    autogobble,
}

\usepackage[normalem]{ulem} 
\newcommand\hlg{\bgroup\markoverwith
  {\textcolor{green!10}{\rule[-.5ex]{2pt}{2.5ex}}}\ULon}
\newcommand\hlb{\bgroup\markoverwith
  {\textcolor{blue!10}{\rule[-.5ex]{2pt}{2.5ex}}}\ULon}
\newcommand\hlr{\bgroup\markoverwith
  {\textcolor{red!10}{\rule[-.5ex]{2pt}{2.5ex}}}\ULon}

\usepackage{listings}

\newif\ifFirstMintedPart


\usepackage[framemethod=TikZ]{mdframed}
\usepackage{tcolorbox}
\makeatletter
\newcommand{\mybox}[1]{%
  \setbox0=\hbox{#1}%
  \setlength{\@tempdima}{\dimexpr\wd0+13pt}%
  \begin{tcolorbox}[boxrule=0.5pt, colback=white, arc=4pt,
      left=6pt,right=6pt,top=6pt,bottom=6pt,boxsep=0pt]
    #1
  \end{tcolorbox}
}
\usepackage{tikz}
\usepackage{pgfplots}
\usetikzlibrary{pgfplots.statistics,calc}
\usepackage{array}
\usepackage{centernot}
\usepackage{xspace}
\usepackage{url}
\usepackage{verbatim}
\usepackage{wrapfig}
\usepackage{tabularx}
\usepackage{bbding}
\usepackage{pifont}
\usepackage{wasysym}
\usepackage{amssymb}

\newcommand{\tool}{\texttt{MemoDroid}}



%% file: sec/introduction.tex
\section{Introduction}
\label{sec_introduction}
Graphical User Interface (GUI) testing is essential for ensuring the quality and reliability of modern mobile applications~\cite{Monkey,yasin2021droidbotx,dong2020time,li2019humanoid,pan2020reinforcement}. Traditional automated GUI testing approaches typically rely on pre-defined exploration scripts or hard-coded strategies, which limit their adaptability to complex and evolving user interfaces. With the rise of Large Language Models (LLMs), researchers and practitioners have begun to build intelligent GUI agents for mobile app testing~\cite{liu2024make,hu2024auitestagent,yoon2023autonomous,ran2024guardian,liu2024vision} and automation~\cite{zhang2023appagent,wang2024mobile,wen2024autodroid,wen2023droidbot}. These GUI agents leverage LLMs' multimodal understanding capabilities to interpret GUI screenshots, reason about app content, and autonomously generate action sequences.

Despite their flexibility and general-purpose reasoning, current LLM-based GUI testing approaches are typically designed for single-session testing, each testing task is executed from scratch, without any memory of prior executions or ability to accumulate targeted experience. While LLMs have world knowledge and reasoning capabilities, they lack specialized, app-specific learning during the testing process. As a result, these LLM-based GUI testing approaches often repeat ineffective interactions and cannot refine their strategies across multiple testing sessions, which limits their testing ability to improve over time. In contrast, human testers continuously build knowledge through hands-on practice, gradually learning to focus on bug-prone functionality, explore untouched areas, and avoid redundant actions.

Many real-world apps share common UI patterns and functional modules. Even within the same app, testers often need to perform testing across multiple versions, updates, or usage scenarios. In such cases, accumulated experience becomes crucial as testers rely on past knowledge to decide which interactions are likely to improve coverage, which functionalities and actions are historically more bug-prone, and which exploration strategies have proven effective in similar contexts. Unlike current LLM-based GUI testing approaches that treat each task as isolated, human testers learn from repeated exposure~\cite{itkonen2015test}, gradually building abstract knowledge of what to explore, what to avoid, and how to adapt. While some recent work explores rule-based memory reuse or external knowledge augmentation~\cite{su2024enhancing,su2022constructing,wang2024knowledge}, these approaches typically depend on heuristic rules or static knowledge bases, requiring manual alignment across app structures and limiting adaptability. There is still a lack of a dynamic, experience-driven memory mechanism that allows LLM-based GUI testing systems to grow over time, reflecting on past behaviors to summarize issue actions, and reusing learned strategies to guide future testing more effectively.

Recent advances across both academia and industry highlight the importance of memory in enabling agents to learn and adapt over time. In language agents~\cite{shao2023character, li2023chatharuhi,wang2023rolellm,zhou2023characterglm,kaiya2023lyfe}, cognitive architectures such as Memochat~\cite{lu2023memochat}, MemGPT~\cite{packer2023memgpt}, and recent frameworks like the Cognitive Architectures for Language Agents (CoALA)~\cite{sumers2023cognitive} demonstrate how 
memory systems support better planning, reflection, and behavior reuse across long-horizon tasks. In domains like code generation~\cite{tsai2024rtlfixer,chen2023gamegpt,li2023metaagents} and robotics~\cite{xie2024embodied,mei2024continuously}, memory has also been shown to enhance agent decision-making by leveraging past traces and outcomes. However, memory design for automated GUI testing remains unexplored and poses domain-specific challenges. Unlike the language tasks, where prior prompts or context can be directly reused, automated GUI testing involves multimodal, dynamic interaction traces that must be abstracted, organized, and matched across heterogeneous app contexts. The memory must not only record low-level test traces but also generalize over issues, summarize functional actions, and offer adaptive guidance in future tasks. 

Inspired by cognitive memory theories that emphasize learning through retention, abstraction, and reuse~\cite{craik1992human,baddeley1997human}, we propose {\tool}, a three-layer memory mechanism for enhancing LLM-based GUI testing. Unlike prior approaches~\cite{li2024appagent,liu2024make,ran2024guardian,hu2024auitestagent} that treat each testing session in isolation, {\tool} accumulates and reuses testing experience across tasks and applications. The \textit{interaction-level episodic memory} segments and encodes functional testing traces, the \textit{function-level reflective memory} summarizes repeated behaviors and identifies failure patterns, and the \textit{app-level strategic memory} synthesizes exploration strategies across apps. During testing, {\tool} dynamically retrieves relevant memory based on current GUI state, stagnation signals, or functional transitions, and injects it into the LLM prompt to guide decision-making. We implement {\tool} as a lightweight plugin that can be integrated into existing LLM-based GUI testing systems without modifying their core architecture. It also allows memory accumulated from previous testing sessions to be reused in future tasks, which enables long-term improvement.

Designing {\tool} involves several key challenges.
First, a challenge lies in how to capture functional-level interaction experience in a structured and reusable form. {\tool} addresses this by segmenting continuous testing traces into discrete functional units based on UI transitions and action semantics. Each unit is represented by annotated screenshots, structured action sequences, and an interaction graph, and is further encoded into a multimodal embedding to support downstream retrieval and analysis. Second, a key question is how to abstract and summarize testing behaviors for effective reuse. {\tool} constructs reflective memory by merging semantically similar function units, analyzing execution patterns such as repeated actions or unresponsive operations, and generating high-level reflections through MLLM-based summarization. These summaries provide a concise and behavior-aware abstraction of past testing outcomes. Third, generalizing testing knowledge across apps requires identifying transferable strategies beyond app-specific traces. {\tool} synthesizes strategic memory by aggregating reflections across apps and prompting the LLM to generate app-level exploration plans. At runtime, {\tool} invokes memory based on GUI similarity and incorporates the retrieved memory into the prompt to inform testing decisions.

We evaluate {\tool} on real-world Android apps from 15 diverse categories. Results show that {\tool} enhances GUI testing performance across five baseline approaches,  with activity and code coverage increasing by 79\% - 96\% and 81\% - 97\%, respectively, and bug detection rate increasing by 57\% - 198\%. An ablation study confirms that all three memory layers contribute to performance gains. Further experiments reveal that expanding the memory pool across 12 rounds leads to progressively better testing outcomes, demonstrating the benefit of experience accumulation. Finally, we conduct usefulness experiments on 200 popular apps on Google Play by integrating baselines with {\tool}, and detect 49 previously unknown crash bugs, 35 of which have been fixed and 14 have been confirmed by developers.

The contributions of this paper are as follows:
\begin{itemize}
\item We propose a three-layer memory mechanism, {\tool}\footnote{We release the source code, dataset, and experimental results on our website \url{https://github.com/TheMystery123/MemoDroid}. \label{github}}, to enhance LLM-based GUI testing, consisting of episodic, reflective, and strategic memory layers that support trace retention, experience summarization, and strategy reuse.

\item We automatically construct and release a structured memory pool covering 60 Android apps\textsuperscript{\ref{github}}, which can be reused to improve future LLM-based GUI testing agents.


\item Effectiveness and usefulness evaluation of the {\tool} in real-world apps with practical bugs detected and confirmed.
\end{itemize}

%% file: sec/motivation.tex
\input{figure/overview}

%% file: figure/overview.tex
\begin{figure*}[htb]
\centering
\includegraphics[width=17.0cm]{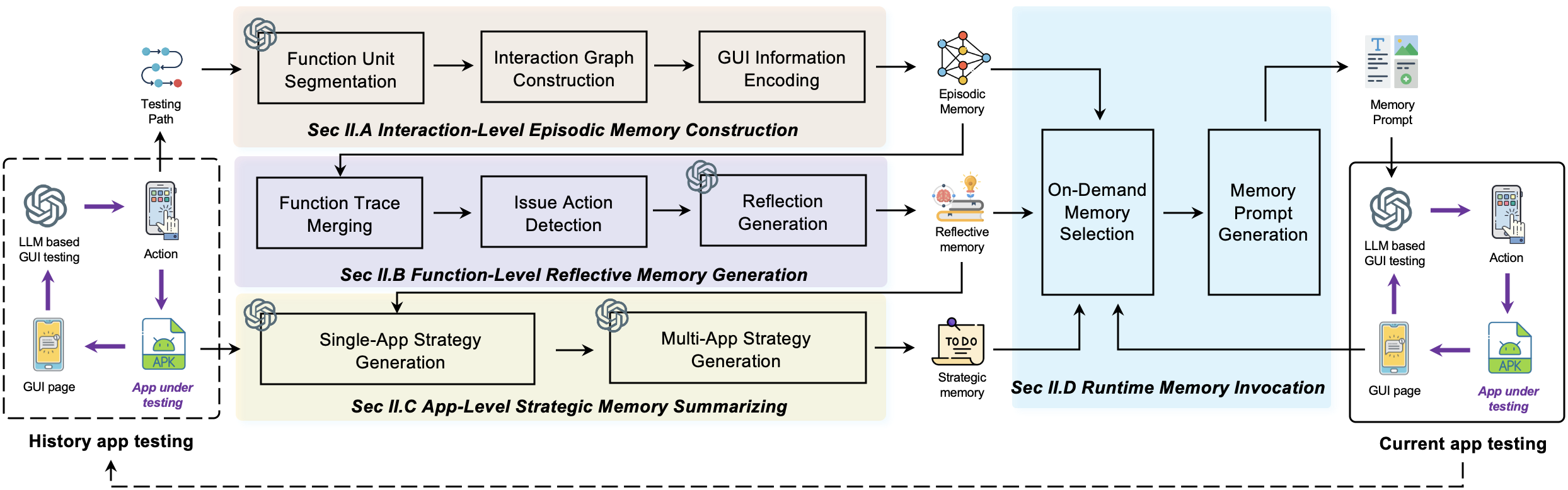}
\caption{Overview of {\tool}}
\label{fig:overview}
\vspace{-0.1in}
\end{figure*}

%% file: sec/approach.tex
\section{Approach}
\label{sec_approach}
This paper proposes {\tool}, a three-layer memory mechanism designed to enhance LLM-based GUI testing by supporting the accumulation, abstraction, and reuse of experience across testing tasks. 
Unlike prior approaches that rely solely on short-term context within a single testing session, {\tool} enables the LLM-based GUI testing (testing agent) to develop long-term, task-agnostic knowledge. 
This memory mechanism allows the agent to retain fine-grained execution traces, summarize testing outcomes, and derive reusable testing strategies, which is similar to how human testers gradually build domain-specific experience through repeated practice.
As shown in Fig.~\ref{fig:overview}, {\tool} structures memory into three complementary layers: \textbf{Interaction-Level Episodic Memory}, \textbf{Function-Level Reflective Memory}, and \textbf{App-Level Strategic Memory}. 
The \textbf{Episodic Memory} records testing traces segmented by functional boundaries, including annotated screenshots, action logs, and GUI transitions. These traces are further encoded using visual-textual information and interaction graph embeddings for downstream reflection. 
The \textbf{Reflective Memory} consolidates repeated executions, identifies issue actions, and summarizes behavioral patterns through structured metadata and natural language generation. 
The \textbf{Strategic Memory} synthesizes multiple reflections not only from a single app but also across apps, producing generalized strategies that highlight priority functionalities, recommended exploration sequences, and bug-prone behaviors. 

{\tool} incorporates a dynamic memory invocation mechanism throughout the testing process. It monitors exploration stagnation and repetitive behavior, then determines when to retrieve relevant memory based on GUI similarity.
Depending on the testing scenario, {\tool} retrieves and injects memory at the start of testing, mid-term testing, or the transition between functionalities. 
{\tool} leverages testing experiences to guide follow-up testing more effectively, plan exploration paths more strategically, and achieve broader coverage while avoiding unnecessary repetition.
We implement {\tool} as a lightweight plugin that can be integrated into existing LLM-based GUI testing systems without modifying their core architecture. 

\subsection{\textbf{Interaction-level Episodic Memory Construction}}
\label{subsec:raw_memory}
The interaction-level episodic memory captures the agent's testing traces at the granularity of app functionalities. 
Unlike prior approaches~\cite{liu2023chatting, ran2024guardian,liu2024vision} that store full-session logs or unstructured action histories, {\tool} segments the testing process into discrete function units, each representing a complete interaction sequence. 
This unit-based organization aligns with how app testing is typically conducted, where human testers also observe, verify, and document app behavior in terms of functionalities rather than isolated actions.

As shown in Fig.~\ref{fig:overview}, the construction of episodic memory involves three components: 
(1) segmenting continuous interaction traces into function units based on GUI transitions and action semantics; 
(2) constructing an interaction graph for each unit, where nodes represent GUI pages and edges represent labeled actions; 
(3) encoding each unit using multimodal representations that integrate visual embeddings, textual interactions, and structural features.

\subsubsection{\textbf{Function Unit Segmentation}}
\label{subsubsec:raw_function_segmentation}
During the automated GUI testing process, the testing agent generates a continuous sequence of GUI states and actions. To organize these traces into episodic memory, we segment them into discrete function units, each representing a self-contained user task such as adding an expense or editing an entry. As shown in Fig.~\ref{fig:overview}, the segmentation consists of three components: 

\textit{Screenshot Annotation.} We first annotate the screenshots captured at each interaction step by identifying the target component using its bounding box from the view hierarchy file. As shown in Fig.~\ref{fig:Annotation} (a), we draw red rectangles over the interactive components using OpenCV's \texttt{rectangle()} function. Each 10 consecutive steps is grouped as a sliding window, and the annotated screenshots are concatenated vertically using OpenCV's \texttt{vconcat()} to form a composite image. Within the image, each action is assigned an index from 1 to 10 for reference.

\textit{Textual Sequence Construction.} For each screenshot in the function unit, we construct a textual sequence describing the interaction as shown in \ref{fig:Annotation} (b). We extract its (1) page index, (2) activity name, (3) action type (`click', `input', `scroll', `long\_click'), and (4) the most informative component identifier, which selected in order of priority from the `text' attribute, then `label', and finally `id'.

\textit{Functional Segmentation with MLLM.} The visual-textual pair is input to a Multimodal Large Language Model (MLLM), which is prompted to analyze the functional structure of the sequence. Rather than outputting a binary decision over the whole window, the MLLM identifies the index ranges within the 10 steps that correspond to complete function units. As shown in \ref{fig:Annotation}, steps 1 to 6 are identified as a coherent function unit corresponding to the process of adding an expense. The generated description is: ``This sequence completes an expense creation process, including amount entry, category selection, and confirmation.'' Once the units within a window are identified, the next window begins at the first unused step. For example, if 1 to 7 is a unit, the next window starts from step 7 and spans pages 7 to 16. This process continues iteratively until the entire trace is segmented.
The output is a list of segmented function units. Each unit consists of: (1) a long image showing the interaction sequence; (2) a structured textual log of the involved steps; and (3) an MLLM-generated natural language summary describing the functionality.

\input{figure/Annotation}

\subsubsection{\textbf{Interaction Graph Construction}}
\label{subsubsec:raw_interaction_graph}
To structurally capture the flow of actions within each function unit, we construct a directed interaction graph  \( G = (V, E) \), where each node \( v_i \in V \) corresponds to a GUI state (i.e., a screenshot and its associated view hierarchy), and each directed edge \( e_{i \rightarrow j} \in E \) represents an interaction transitioning the app from \( v_i \) to \( v_j \).
Each edge is labeled with the action (e.g., click, scroll, input) and the corresponding component identifier (component's text, label, or ID). This labeling preserves both the transition semantics and component-level behavior. The interaction graph is later used during the embedding phase.

\subsubsection{\textbf{GUI Information Encoding}}
\label{subsubsec:raw_embedding_encoding}
To enable memory retrieval and similarity comparison across function units, each segmented unit is encoded into a compact multimodal representation. 
We adopt a dual-encoder architecture based on the CLIP model~\cite{Radford2021LearningTV}. For each function unit, the annotated screenshots are resized to a fixed resolution and individually passed through a pre-trained CLIP ViT-B/16 vision encoder, yielding one visual embedding per step. These embeddings are then aggregated via average pooling to obtain a single visual vector for the unit.
In parallel, we encode the textual sequence of the function unit using the CLIP text encoder, yielding a global textual embedding that summarizes the interaction history in natural language.
To incorporate behavioral structure, we apply a two-layer Graph Convolutional Network (GCN) over the interaction graph described in Section~\ref{subsubsec:raw_interaction_graph}. 
Finally, we concatenate the visual, textual, and structural embeddings and project the result into a unified latent space using a fully connected layer with ReLU activation. The resulting vector is stored in episodic memory and used for future retrieval, clustering, and similarity alignment.
In addition to encoding full function units, we also apply the same procedure to encode the first screenshot of each unit independently. This enables finer-grained retrieval in Section \ref{sec:memory_invocation}, especially when only the first page of a function is visible at the beginning of new functions.


\input{figure/FMemory}

\subsection{\textbf{Function-Level Reflective Memory Generation}}
\label{subsec:reflective_memory}
To enable experience abstraction and facilitate high-level memory reuse, {\tool} constructs a function-level reflective memory by analyzing episodic memory and summarizing its behavioral characteristics. 
{\tool} first merges functionally similar testing traces to reduce redundancy.
It then identifies issue-triggering actions during testing, which are key interactions likely to reveal issues or abnormal behaviors. 
{\tool} finally generates a natural language summary for each function unit by prompting an MLLM with the annotated screenshots and structured metadata from the function units.

\subsubsection{\textbf{Function Trace Merging}}
\label{subsubsec:reflect_merge}
{\tool} merges functionally similar units based on three aspects: the activity name sequence, the sequence of action-component pairs, and the GUI page structure across steps. 
For each function unit, {\tool} extracts: (1) the activity name at each step; (2) the action type with its corresponding UI component identifier (text, label, or ID); and (3) a structure hash computed from the view hierarchy of each GUI page. We compute a weighted similarity score by combining the Levenshtein similarity of activity sequences, the Jaccard similarity of action-component sets, and the perceptual hash match score between GUI pages. Scores that exceed a predefined threshold of 0.8 (Based on our pilot study and existing work~\cite{pan2020reinforcement,liu2022Guided,su2017guided}) are considered equivalent and merged.
After merging, we construct a JSON object to summarize the behavior across merged traces. This includes: the number of times the function was executed, the average path length, the number of unique screens, duplicate screen counts, screen occurrence distribution, the most frequent start page, and the functional description inherited from episodic memory.

\subsubsection{\textbf{Issue Action Detection}}
\label{subsubsec:reflect_issue_detection}
{\tool} identifies and records issue actions within each function unit. These issue actions are stored as part of the reflective memory and provide actionable insights for strategy generation and bug identification.
We detect three types of issue actions. For each, we record the associated component and screen to support traceability and related analysis.
First, \textit{unresponsive actions}, if the GUI remains unchanged after interaction. We compare the pre- and post-action screenshots and view hierarchy trees; if no visible or structural difference is found, the operation is classified as unresponsive.
Second, \textit{redundant actions}, when the same action is repeatedly executed on the same component across multiple steps. These patterns often indicate exploration inefficiencies or unintended loops. For each such case, we identify
the component, its screen, and repetition count.
Third, \textit{bug-triggering actions}, if they cause a bug such as a crash. The triggering component and screen are recorded for each instance.
All detected issue actions are embedded into the structured metadata of the reflective memory and used to generate natural language reflections and inform strategy-level planning.

\subsubsection{\textbf{Natural Language Reflection Generation}}
\label{subsubsec:reflect_summary}
{\tool} generates a high-level natural language reflection for each function unit, which summarizes the testing behavior and issue actions.
As shown in Fig. \ref{fig:FMemory}, the input consists of: (1) the image of the function unit; and (2) a \textit{testing behavior summary},
which records testing behavior data aggregated during the merging (Section \ref{subsubsec:reflect_merge}) and issue action detection (Section \ref{subsubsec:reflect_issue_detection}).

We construct a prompt that incorporates both the visual and structured information. This prompt is input for an MLLM, which is instructed to generate a concise reflection describing the observed issues, behavior summary, cause, and suggestions as shown in Fig. \ref{fig:FMemory} (c). 
These reflections provide both human-readable insights and LLM-compatible summaries that can be recalled during future tests. 

\subsection{\textbf{App-Level Strategic Memory Summarizing}}
\label{subsec:strategic_memory}
To support planning at the app level and facilitate experience transfer across applications, {\tool} constructs a strategic memory layer that stores high-level testing strategies derived from reflective memory. These strategies summarize functional testing priorities, recurrent interaction risks, and bug-trigger actions commonly observed in specific app categories or usage scenarios.
We construct a prompt that integrates the app metadata and the associated reflections. This prompt is provided to an MLLM, which is instructed to generate an app-level strategy summary. 

\subsubsection{\textbf{App-Level Strategy Summary Generation}}
\label{subsubsec:strategic_app_summary}
The input includes: (1) the app's name, category, and description; and (2) the set of natural language reflections.
As shown in Fig.~\ref{fig:SMemory} (a), we integrate these inputs into a structured prompt, it ends with a query requesting a concise and reusable testing strategy.
As shown in Fig.~\ref{fig:SMemory} (b), the output includes three parts: (1) a global testing plan indicating testing priorities and suggested execution order; (2) function-level strategies offering specific testing for key operations; and (3) actions that may trigger bugs.
The generated strategy is stored in textual format as part of the app's strategic memory. It serves as reusable guidance when retesting the same apps or testing apps with similar structures or functionality.

\input{figure/SMemory}

\subsubsection{\textbf{Cross-App Strategy Abstraction}}
\label{subsubsec:cross_app_strategy}
To further enhance the generalizability of testing knowledge, we abstract higher-level strategies by aggregating the app-specific summaries across multiple apps belonging to the same category, such as e-commerce, social media, or utilities. The input to this step consists of the natural language strategy summaries generated for each app in the previous stage. Each summary encapsulates the testing experiences and behavioral observations specific to one app. We group these summaries based on app categories. For each category, we concatenate the app-level strategy summaries and use an MLLM to synthesize a consolidated domain-level testing strategy, which is instructed to identify recurring patterns, frequent sources of bugs, and best practices that apply across the set of applications.

\subsection{\textbf{Runtime Memory Invocation}}
\label{sec:memory_invocation}
Runtime memory invocation refers to the process by which {\tool} dynamically retrieves relevant memory entries during testing. This mechanism enables the system to leverage accumulated experience to guide decision-making in contextually appropriate moments. {\tool} triggers memory invocation at the start of a new app, testing stagnation, or functional transitions. It analyzes the current GUI state, recent interaction history, and overall app metadata to retrieve related memory from the episodic, reflective, and strategic memory. Retrieved content is organized into a prompt, which is integrated into the LLM input to enhance its reasoning and action selection, without modifying the underlying testing framework.

\subsubsection{\textbf{On-Demand Memory Selection}}
\label{subsec:memory_trigger_matching}
To enable memory utilization during testing, {\tool} adopts an on-demand memory selection mechanism, where memory is retrieved only under specific runtime conditions. This selective strategy avoids unnecessary overhead while ensuring that guidance is provided when it is most impactful. {\tool} defines three types of triggers: cold start, testing stagnation, and functional transitions.


\textit{Cold Start.} At the beginning of testing a new app, {\tool} retrieves app-level strategies from strategic memory. The retrieval is based on app textual information, including the app name, category, app description, and the list of activity names. These inputs are encoded into sentence embeddings and compared the app textual information of existing strategic memory entries using cosine similarity. 

\textit{Mid-testing Stagnation.} During testing, {\tool} monitors the testing trace and detects stagnation based on: (1) more than 60\% of recent actions revisit previously explored screens; (2) fewer than 3 unique action types appear in the past 10 steps; (3) no new function unit is completed within the last 20 steps. When stagnation is detected, {\tool} queries the reflective memory using the encoding of the most recent function unit in Section \ref{subsubsec:raw_embedding_encoding}. This encoding includes visual, textual, and graph structure information of the function unit.

\textit{Functional Transitions.} After completing one function unit, {\tool} anticipates a transition to a new functionality. It retrieves episodic memory entries related to the current GUI page, including the current page's screenshot and textual description, encoded using the same vision-language encoder as in Section~\ref{subsubsec:raw_embedding_encoding}. These embeddings are matched against the first-page embeddings of previously tested function units in memory. The retrieved episodic memory contains natural language summaries describing what functionality was previously explored from the similar UI page, including functionality description and the action sequences.


\subsubsection{\textbf{Memory Prompt Generation}}
\label{subsec:prompt_injection}
After retrieving relevant memory, {\tool} organizes them into structured prompts and injects them into the front of the prompt as the LLM's input of LLM-based GUI testing to guide testing decisions.

When strategic memories are retrieved during cold start, the app-level strategic memory are inserted into the prompt as high-level exploration guidance. Memory is prefixed with a system instruction indicating it originates from prior testing of similar apps and should be considered when planning initial actions. During testing stagnation, reflective memory are injected with the prompt to provide behavioral references, which describe the execution trace of a similar function unit. During functional transitions, {\tool} retrieves function units from episodic memory whose starting pages match the current pages. The corresponding function unit testing tace are inserted into the prompt.


\subsection{\textbf{Implementation}}
\label{subsec_approach_Implement}
In our implementation, we utilize the GPT-4o-mini\footnote{\url{https://openai.com/index/gpt-4o-mini-advancing-cost-efficient-intelligence/}} to handle both textual and visual information during the testing process. 
It obtains the view hierarchy file of the current GUI page through UIAutomator~\cite{uiautomator} to extract text information of the input widgets.
{\tool} is implemented as a lightweight plugin, which is compatible with existing LLM-based GUI testing systems and can be integrated without modifying their core architecture.

%% file: figure/Annotation.tex
\begin{figure*}[htb]
\vspace{0.1in}
\centering
\includegraphics[width=17.0cm]{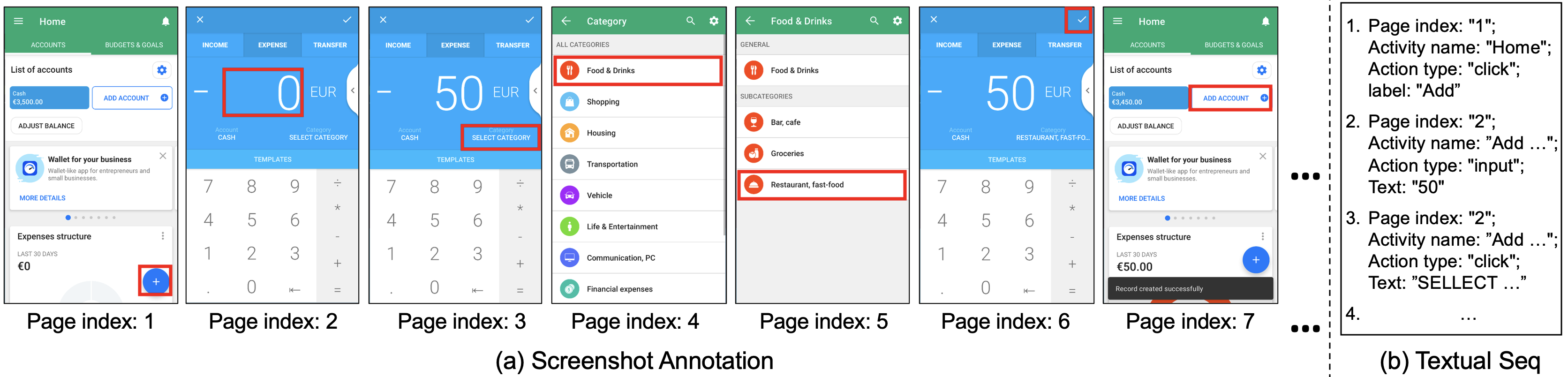}
\caption{Example of {\tool}'s screenshot annotation and textual sequence construction}
\label{fig:Annotation}
\vspace{-0.1in}
\end{figure*}

%% file: figure/FMemory.tex
\begin{figure*}[htb]
\vspace{-0.1in}
\centering
\includegraphics[width=17.0cm]{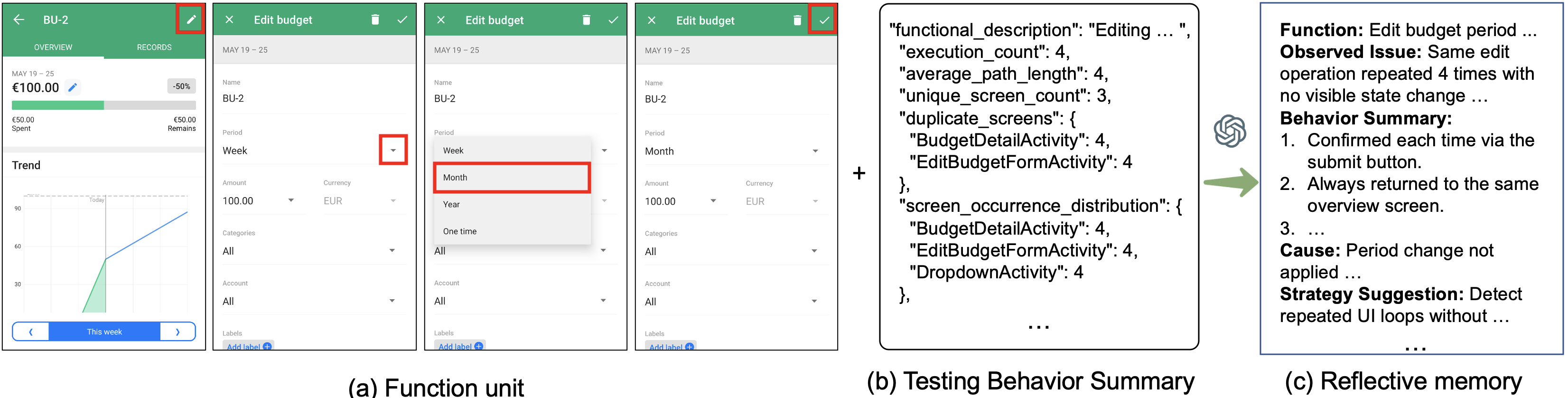}
\caption{Example of reflective memory generation}
\label{fig:FMemory}
\vspace{-0.1in}
\end{figure*}

%% file: figure/SMemory.tex
\begin{figure}
\centering
\includegraphics[width=8.3cm]{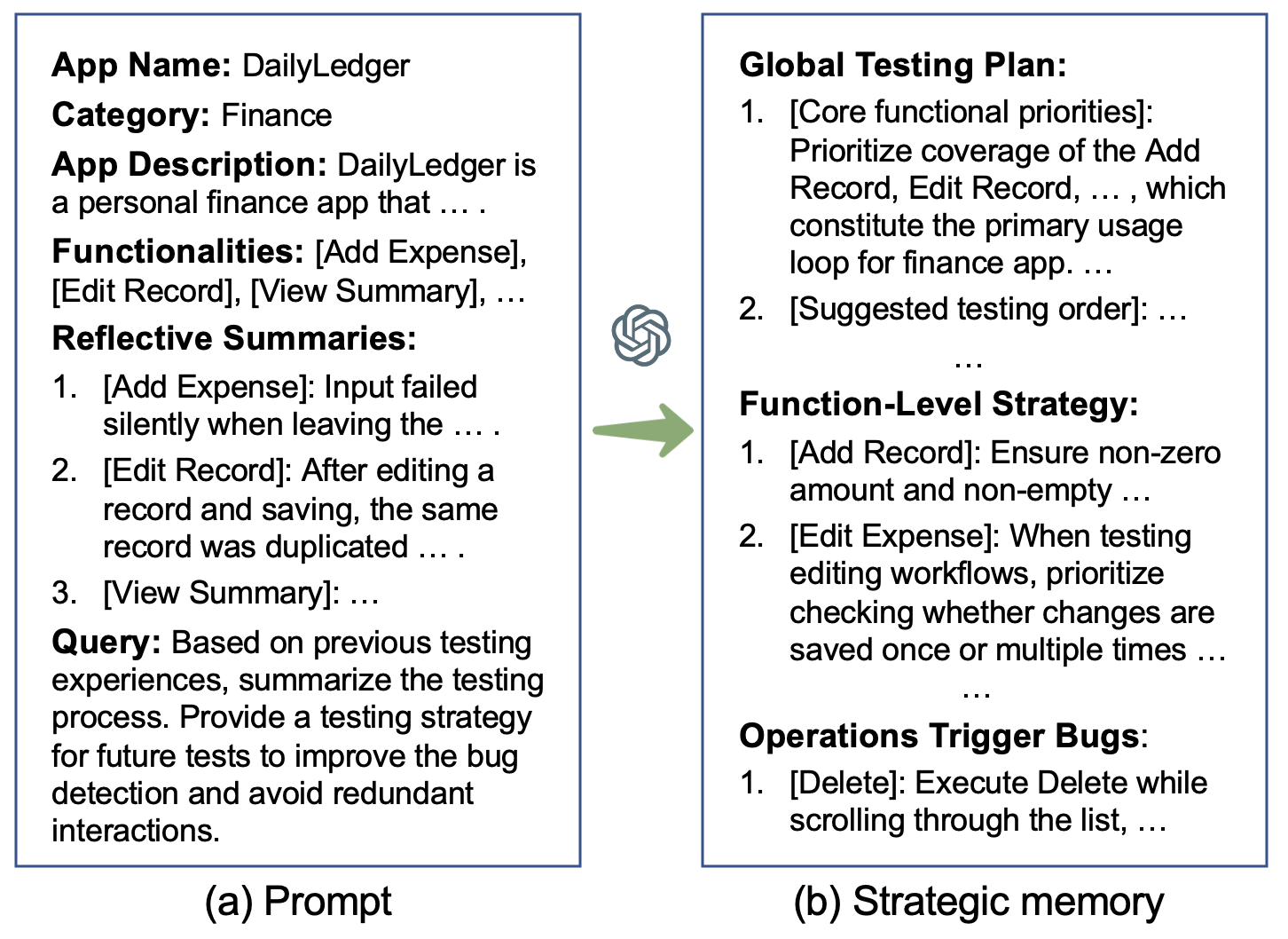}
\caption{Example of strategic memory generation}
\label{fig:SMemory}
\vspace{-0.1in}
\end{figure}

%% file: sec/experiment.tex
\section{Experiment Design}
\label{sec_Experiment_Design}

We design four research questions to evaluate the effectiveness, component contribution, and practical usefulness of {\tool}:

\begin{itemize}[leftmargin=*]
\item \textbf{RQ1 (Effectiveness Evaluation):} Can memory accumulated from previously tested apps improve GUI testing performance on new apps across categories?
\item \textbf{RQ2 (Ablation Study):} How does each memory layer (episodic, reflective, strategic) contribute to the overall effectiveness of {\tool}?
\item \textbf{RQ3 (Memory Evolving
Effect):} Does expanding the memory pool continuously improve GUI testing performance?
\item \textbf{RQ4 (Usefulness Evaluation):} Can LLM-based GUI testing with {\tool} discover unknown bugs in large-scale real-world apps?
\end{itemize}







\subsection{\textbf{Dataset}}
\label{subsec_experiment_dataset}

\textbf{For RQ1 and RQ2}, we evaluate whether testing memory from previously explored apps can enhance GUI testing performance on new apps. The experimental data consists of two parts: a \textit{memory pool} and a \textit{testing set}. 

We do not adopt existing benchmarks such as Themis~\cite{su2021benchmarking} due to their limited scale (20 apps) and outdated content (mostly before 2019), which are often incompatible with current Android systems. Following the collection principles of Themis, we curate our own dataset from F-Droid by selecting apps that: (1) were updated after January 2025, (2) are compatible with all baseline methods, (3) contain at least 12 activities, (4) support UIAutomator-based screenshot and view hierarchy extraction, and (5) include at least one manually verifiable functional bug. 

We collect 90 apps that satisfy these criteria, covering 15 distinct categories with 6 apps per category. Among them, 60 apps (4 per category) are used to build the memory pool, while the remaining 30 apps (2 per category) form the testing set. The testing set apps are strictly excluded from memory construction to ensure fair evaluation.

\textbf{For RQ3}, we study whether an evolving memory pool leads to continuous improvements in testing performance. 
Specifically, we randomly partition the 60 apps used for memory construction into 12 groups, each containing 5 apps. These groups are incrementally added to the memory pool across 12 rounds. In round $i$, the memory pool consists of the first $i$ apps. After each round, we test on the same 30 apps from RQ1 to measure how memory evolution affects generalization.

\textbf{For RQ4}, we assess the real-world usefulness of {\tool} by applying it to 200 popular Android apps from Google Play, covering a diverse range of categories. These apps are selected based on popularity and recent update time. We initialize the memory with the full set of 60 apps from RQ1 and evaluate whether memory-guided testing can uncover previously unknown bugs. Once potential issues are found, we report them to developers through GitHub or official email channels and track whether the bugs are confirmed or fixed.

\subsection{\textbf{Baselines}}
\label{subsec_experiment_baseline}
To evaluate the effectiveness of our proposed memory mechanism, we compare {\tool} with five state-of-the-art LLM-based GUI testing approaches that do not incorporate memory support. Previous studies shows that these LLM-based methods are more effective than traditional GUI testing methods. However, we do not compare them with traditional methods again.
\textbf{GPTDroid}~\cite{liu2024make} transforms GUI testing into a question-answering process with LLMs. It encodes GUI state into structured prompts, elicits next actions from the LLM.
\textbf{DroidAgent}~\cite{yoon2023autonomous} introduces an intent-driven multi-agent framework for LLM-based testing. It generates semantic task goals and decomposes them into actionable steps.
\textbf{AUITestAgent}~\cite{hu2024auitestagent} combines dynamic proxy organization and multi-dimensional GUI information extraction. 
\textbf{VisionDroid}~\cite{liu2024vision} employs a multi-agent MLLM framework. It aligns GUI screenshots with textual information and coordinates three agents to generate the testing path.
\textbf{Guardian}~\cite{ran2024guardian} enhances LLM-based GUI testing through external runtime enforcement. It optimizes the action space using domain constraints and supports dynamic re-planning by offloading inconsistent test paths, improving the stability and focus of LLM decisions.
All baselines are evaluated using their released implementations or replicated pipelines following original configurations. 

\subsection{\textbf{Experimental Setup}}
\label{subsec_experiment_Setup}
We implement {\tool} as a lightweight plugin module integrated into existing LLM-based GUI testing frameworks. All experiments are conducted on a workstation equipped with an Intel Core i7-12700F CPU, 32GB DDR5 RAM, and a single NVIDIA GeForce RTX 3060 GPU (12GB), running Android emulator instances to ensure consistency across runs.
For each experiment, we compare the performance of baseline testing methods with and without {\tool} augmentation. To ensure fairness, both versions use the same LLM backbone and are initialized with identical configurations. Each tool is given 60 minutes to test each app, following the common practice in prior studies~\cite{gu2019practical,su2021benchmarking,li2017droidbot}. For apps requiring login, we register test accounts and pre-script login flows. Before each run, app data is cleared to avoid cross-run interference.
We run each tool three times and obtain the average performance to mitigate potential bias.

To construct the memory pool used in RQ1, RQ2, and RQ4, we first run five LLM-based GUI testing baselines on 60 apps (Section~\ref{subsec_experiment_dataset}). Each app is tested independently, and the testing traces are recorded and organized into memory following the layered structure of {\tool}. These traces are later encoded into episodic, reflective, and strategic memory components to enable experience reuse during testing of unseen apps. For each baseline, the same memory pool is used for fair comparison against its {\tool} enhanced variant.

For RQ3, we begin with an empty memory and incrementally grow the pool by adding one app at a time from a preselected set of 60 apps. In each of the 12 rounds, the memory pool consists of the first $i$ groups (i.e., $5 \times i$ apps). After each update, we evaluate testing performance on the same 30 apps used in RQ1. 



\subsection{\textbf{Evaluation Metrics}}
We measure the performance from bug detection and testing coverage of our {\tool} and baselines.
For bug detection, we use the number of bugs detected by automated GUI testing tools. 
For testing coverage, we obtain the activity coverage and code coverage~\cite{he2020textexerciser,liu2017automatic,arnatovich2018mobolic,wang2020combodroid,wang2021vet,li2019humanoid}, in which we treat the activities defined in the \textit{AndroidManifest.xml} file of an Android app as the whole set of  activities~\cite{pan2020reinforcement,Android,su2017guided}.

%% file: sec/result.tex
\section{Results and Analysis}
\label{sec_result}
\subsection{\textbf{Effectiveness of GUI Testing with {\tool} (RQ1)}}
\label{sec:results_rq1}
Table~\ref{tab:RQ2} shows the average testing performance of five baseline approaches, and they are integrated with {\tool} across 30 testing apps. 
We observe that integrating with {\tool} improves the performance of all baselines. Compared to their original versions, all memory-enhanced variants achieve higher bug detection, activity and code coverage.

\input{tab/RQ2}

\input{figure/RQ3-1}

Specifically, activity coverage increases by 79\% - 96\%, and code coverage improves by 81\% - 97\%, indicating that {\tool} enables more effective activity and code exploration. In terms of bug detection, the number of bugs discovered increases by 57\% - 198\%. These results show that LLM-based GUI testing with {\tool} not only expands coverage but also improves the ability to focus on bug-prone behaviors. The performance gains hold across both text-based and vision-based baselines, confirming the general applicability of {\tool}.

We further manually review the testing records and analyze the reasons for the performance improvement of the 5 GUI testing approaches integrated with {\tool}. First, the \textit{episodic memory} provides low-level trace reuse, which allows the GUI testing approaches to avoid redundant exploration and replicate effective actions. Second, the \textit{reflective memory} identifies issue actions. For instance, It helps GUI testing approaches understand that submitting forms with empty fields or switching certain settings often leads to unexpected crashes.
Third, the \textit{strategic memory} provides high-level exploration strategies to assist GUI testing approaches in making decisions in unfamiliar applications.

\subsection{\textbf{Ablation Study (RQ2)}}
\label{sec:results_rq2}
To evaluate the contribution of each memory layer in {\tool}, we conduct an ablation study by selectively removing one layer at a time and comparing the performance with the full system. Specifically, we consider three variants: \textit{w/o Episodic Memory}, \textit{w/o Reflective Memory}, and \textit{w/o Strategic Memory}, which delete memory from different layers. 

\input{tab/RQ3}

As shown in Table~\ref{tab:RQ3}, removing any memory layer leads to a notable drop in performance, confirming that all three components are essential. Among them, reflective memory contributes the most. Removing it results in a 38\% - 25\% reduction in activity coverage, 37\% - 25\% in code coverage, and 39\% - 19\% fewer bugs detected. This highlights the importance of behavioral summarization and failure-driven guidance.
Episodic memory also plays a key role by supporting low-level reuse. Its removal causes a 26\% - 23\% drop in activity coverage and a 29\% - 22\% drop in code coverage. Strategic memory contributes high-level testing guidance and yields a 22\% - 15\% reduction in activity coverage and a 22\% - 16\% drop in code coverage when removed.

Overall, these results demonstrate that {\tool} achieves its effectiveness by combining all three types of memory: reflective memory offers the strongest improvement, while episodic and strategic memory provide complementary support for trace reuse and test planning.

\subsection{\textbf{Impact of Memory Pool Evolving (RQ3)}}
\label{sec:results_rq3}
To evaluate whether the evolution of memory pools will gradually improve the performance of testing, we conduct a memory evolution experiment. Fig.~\ref{fig:RQ3-1} shows the results across five baselines. We observe an upward trend across all three metrics as the memory pool evolves. This demonstrates that {\tool} benefits from accumulated experience even when the memory is constructed from a small number of diverse apps. After about 5 apps, the baselines with {\tool} begin to outperform them without {\tool}. As more apps are added, the performance gap widens, confirming that {\tool} can generalize across categories.

For instance, \textit{VisionDroid + {\tool}} improves its activity coverage from 0.41 (baseline) to 0.71 after 20 memory apps, while bug detection increases from 1.23 to 1.90. \textit{Guardian + {\tool}} shows a similar trend, with code coverage rising from 0.38 to 0.69. The improvement is smooth,
indicating that {\tool} can continually integrate new testing memory and experience to support long-term testing performance growth.
These results validate the scalability of {\tool}'s memory mechanism. It not only enables individual reuse of prior testing history but also accumulates testing strategies and behavior patterns in a way that generalizes across unseen apps.

\input{tab/RQ4}

\subsection{\textbf{Usefulness Evaluation (RQ4)}}
\label{sec_results_RQ4}
We follow the settings in RQ1 in Section \ref{sec:results_rq1} and run both the original baseline approaches and the baseline approaches integrated with our {\tool} to test 200 apps.
For the 200 apps, {\tool} detects 93 crash bugs in 66 apps, of which 49 bugs in 47 apps are never discovered before. 
Furthermore, as shown in Table \ref{tab:RQ4}, only 7 of these new bugs are detected by the best baseline without {\tool}, and the bugs detected by these baselines are all subsets of the baseline with {\tool}.
We submitted these 49 bugs to developers, and all of them have been fixed/confirmed so far (35 fixed and 14 confirmed, none of them rejected). 
This further demonstrates the effectiveness of our proposed {\tool} in assisting automated GUI testing approaches in detecting bugs.
Due to space limitations, Table \ref{tab:RQ4} presents a part of fixed/confirmed bugs, and the full lists can be found on our website\textsuperscript{\ref{github}}.
We also collect the feedback from developers regarding the bugs we submitted.``This bug affects a core function and somehow slipped through our internal testing.'', ``We weren't aware of this bug.'' These feedbacks also demonstrate the importance of {\tool} in assisting automated GUI testing approaches.

%% file: tab/RQ2.tex
\begin{table}[h]
\vspace{-0.05in}
\caption{Comparison between {\tool} and baselines.}
\renewcommand\arraystretch{1.1} 
\label{tab:RQ2}
\centering
\footnotesize
\begin{tabular}{p{2.5cm}<{\centering} | p{1.4cm}<{\centering} | p{1.4cm}<{\centering} | p{1.6cm}<{\centering}}
\toprule
\multirow{2}*{\textbf{GUI Testing Method}} & \multicolumn{2}{c|}{\textbf{Average Coverage}} & \multirow{2}*{\textbf{Average Bugs}} \cr 
& \textbf{Activity} & \textbf{Code} & \\ 
\midrule
GPTDroid & 0.29  & 0.27  & 0.50 \\ 
\rowcolor{gray!15}
\textbf{GP + {\tool}} & \textbf{0.52 $\uparrow$ 79\%}  & \textbf{0.49 $\uparrow$ 81\%}  & \textbf{1.47 $\uparrow$ 194\%} \\ 
\midrule
DroidAgent & 0.28  & 0.27  & 0.57 \\ 
\rowcolor{gray!15}
\textbf{DA + {\tool}} & \textbf{0.55 $\uparrow$ 96\%}  & \textbf{0.51 $\uparrow$ 89\%}  & \textbf{1.70 $\uparrow$ 198\%} \\ 
\midrule
AUITestAgent & 0.37  & 0.36  & 1.10 \\ 
\rowcolor{gray!15}
\textbf{ATA + {\tool}} & \textbf{0.69 $\uparrow$ 86\%}  & \textbf{0.67 $\uparrow$ 86\%}  & \textbf{1.73 $\uparrow$ 57\%} \\ 
\midrule
VisionDroid & 0.41  & 0.39  & 1.23 \\ 
\rowcolor{gray!15}
\textbf{VD + {\tool}} & \textbf{0.77 $\uparrow$ 88\%}  & \textbf{0.73 $\uparrow$ 87\%}  & \textbf{2.10 $\uparrow$ 71\%} \\ 
\midrule
Guardian & 0.40  & 0.38  & 1.30 \\ 
\rowcolor{gray!15}
\textbf{GU + {\tool}} & \textbf{0.77 $\uparrow$ 93\%}  & \textbf{0.75 $\uparrow$ 97\%}  & \textbf{2.23 $\uparrow$ 72\%} \\ 
\bottomrule
\end{tabular}
\vspace{0.05in}
\begin{tablenotes}
\footnotesize
\item \textbf{\textit{Notes:}} ``GP'' is the GPTDroid, ``DA'' is the DroidAgent, ``ATA'' is the AUITestAgent, ``VD'' is the VisionDroid, ``GU'' is the Guardian.
\end{tablenotes}
\vspace{-0.05in}
\end{table}

%% file: figure/RQ3-1.tex
\begin{figure*}[htb]
\vspace{-0.1in}
\centering
\includegraphics[width=16.5cm]{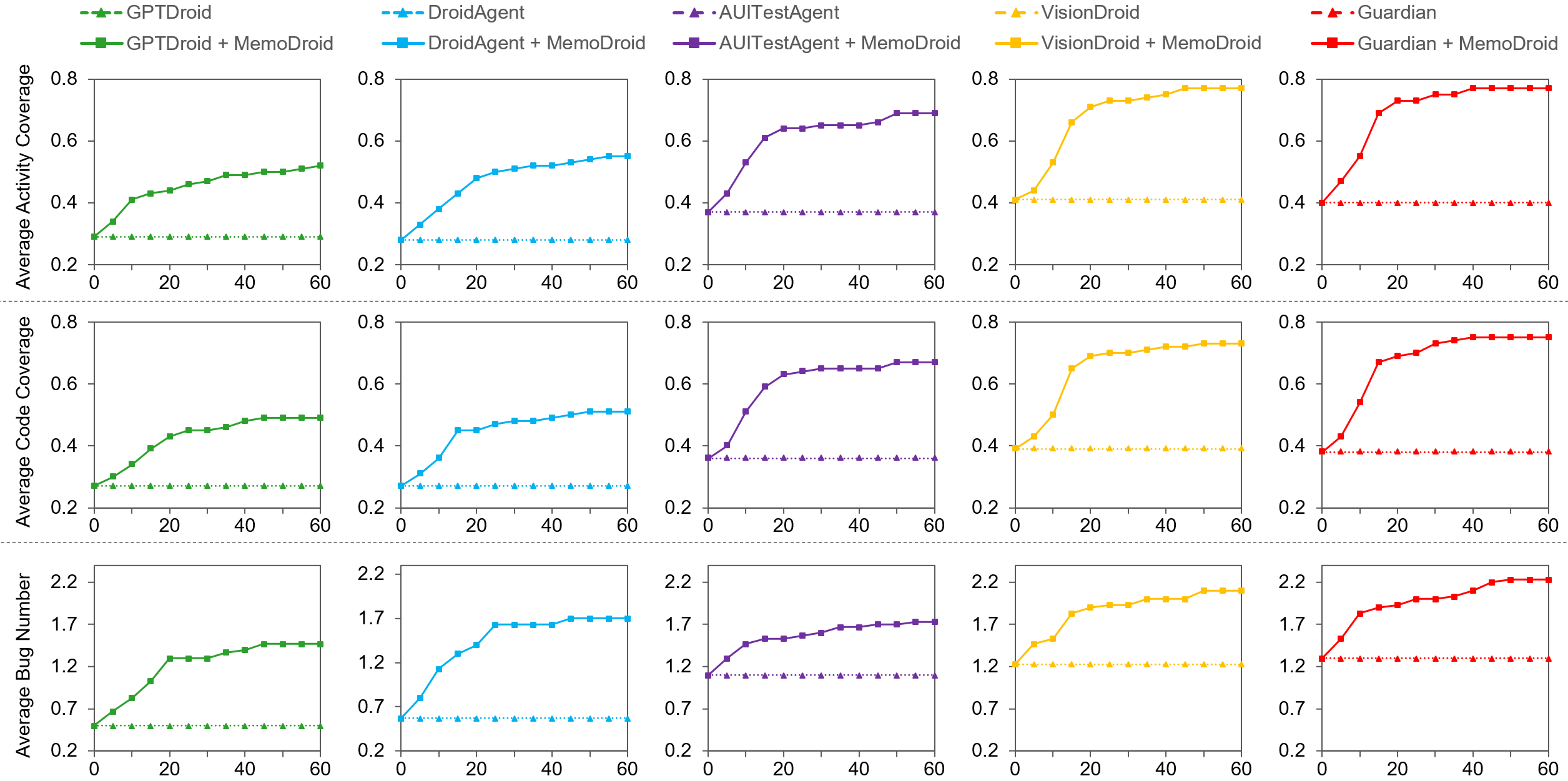}
\caption{Impact of memory pool evolving}
\label{fig:RQ3-1}
\vspace{-0.1in}
\end{figure*}

%% file: tab/RQ3.tex
\begin{table}[h]
\vspace{0.05in}
\caption{Comparison between {\tool} and variants.}
\renewcommand\arraystretch{1.1} 
\label{tab:RQ3}
\centering
\footnotesize
\begin{tabular}{p{2.7cm}<{\centering} | p{1.4cm}<{\centering} | p{1.4cm}<{\centering} | p{1.6cm}<{\centering}}
\toprule
\multirow{2}*{\textbf{GUI Testing Method}} & \multicolumn{2}{c|}{\textbf{Average Coverage}} & \multirow{2}*{\textbf{Average Bugs}} \cr 
& \textbf{Activity} & \textbf{Code} & \\ 
\midrule
\textbf{GP + {\tool}} & \textbf{0.52 }  & \textbf{0.49}  & \textbf{1.47} \\ 
\rowcolor{gray!15}
w/o Episodic Memory  & 0.39 $\downarrow$ 25\% & 0.38 $\downarrow$ 22\% & 1.13 $\downarrow$ 23\% \\ 
w/o Reflective Memory  & 0.36 $\downarrow$ 31\% & 0.34 $\downarrow$ 31\% & 1.00 $\downarrow$ 32\% \\ 
\rowcolor{gray!15}
w/o Strategic Memory  & 0.44 $\downarrow$ 15\% & 0.41 $\downarrow$ 16\%  & 1.23 $\downarrow$ 16\% \\ 
\midrule

\textbf{DA + {\tool}} & \textbf{0.55}  & \textbf{0.51}  & \textbf{1.70} \\ 
\rowcolor{gray!15}
w/o Episodic Memory  & 0.41 $\downarrow$ 25\% & 0.39 $\downarrow$ 24\% & 1.30 $\downarrow$ 24\% \\ 
w/o Reflective Memory  & 0.34 $\downarrow$ 38\% & 0.32 $\downarrow$ 37\% & 1.03 $\downarrow$ 39\% \\ 
\rowcolor{gray!15}
w/o Strategic Memory  & 0.43 $\downarrow$ 22\% & 0.41 $\downarrow$ 20\% & 1.37 $\downarrow$ 19\% \\ 
\midrule

\textbf{ATA + {\tool}} & \textbf{0.69}  & \textbf{0.67}  & \textbf{1.73} \\ 
\rowcolor{gray!15}
w/o Episodic Memory  & 0.52 $\downarrow$ 25\% & 0.51 $\downarrow$ 24\% & 1.40 $\downarrow$ 19\% \\ 
w/o Reflective Memory  & 0.49 $\downarrow$ 29\% & 0.47 $\downarrow$ 30\% & 1.27 $\downarrow$ 27\% \\ 
\rowcolor{gray!15}
w/o Strategic Memory  & 0.55 $\downarrow$ 20\% & 0.53 $\downarrow$ 21\% & 1.50 $\downarrow$ 13\% \\ 
\midrule

\textbf{VD + {\tool}} & \textbf{0.77}  & \textbf{0.73}  & \textbf{2.10} \\ 
\rowcolor{gray!15}
w/o Episodic Memory  & 0.57 $\downarrow$ 26\% & 0.52 $\downarrow$ 29\% & 1.80 $\downarrow$ 14\% \\ 
w/o Reflective Memory  & 0.53 $\downarrow$ 31\% & 0.49 $\downarrow$ 33\% & 1.70 $\downarrow$ 19\% \\ 
\rowcolor{gray!15}
w/o Strategic Memory  & 0.61 $\downarrow$ 21\% & 0.57 $\downarrow$ 22\% & 1.87 $\downarrow$ 11\% \\ 
\midrule

\textbf{GU + {\tool}} & \textbf{0.77}  & \textbf{0.75}  & \textbf{2.23} \\ 
\rowcolor{gray!15}
w/o Episodic Memory  & 0.59 $\downarrow$ 23\% & 0.57 $\downarrow$ 24\% & 1.77 $\downarrow$ 21\% \\ 
w/o Reflective Memory  & 0.58 $\downarrow$ 25\% & 0.56 $\downarrow$ 25\% & 1.60 $\downarrow$ 28\% \\ 
\rowcolor{gray!15}
w/o Strategic Memory  & 0.62 $\downarrow$ 19\% & 0.59 $\downarrow$ 21\% & 1.83 $\downarrow$ 18\% \\ 
\bottomrule
\end{tabular}
\vspace{0.05in}
\begin{tablenotes}
\footnotesize
\item \textbf{\textit{Notes:}} ``GP'' is the GPTDroid, ``DA'' is the DroidAgent, ``ATA'' is the AUITestAgent, ``VD'' is the VisionDroid, ``GU'' is the Guardian.
\end{tablenotes}
\vspace{-0.15in}
\end{table}

%% file: tab/RQ4.tex
\begin{table}[htb]
\vspace{0.05in}
\renewcommand\arraystretch{0.95} 
\caption{Information of the fixed bugs.}
\label{tab:RQ4}
\centering
\scriptsize
\begin{tabular}{p{0.2cm}<{\centering} | p{1.1cm}<{\centering} | p{0.8cm}<{\centering} | p{0.5cm}<{\centering} | p{0.7cm}<{\centering} | p{0.2cm}<{\centering} | p{0.2cm}<{\centering} | p{0.3cm}<{\centering} | p{0.2cm}<{\centering} | p{0.2cm}<{\centering}}
\toprule
\textbf{ID} & \textbf{App name} & \textbf{Category} & \textbf{Down} & \textbf{Version} & \textbf{GP} & \textbf{DA} & \textbf{ATA} & \textbf{VD} & \textbf{GU}\\
 \midrule
1 & Traveloka & Travel & 50M+ & 5.20.0 &  &  &  &  & \\
\rowcolor{gray!15}
2 & Yelp & Food & 50M+ & 25.11 &  &  &  &  & \\
3 & BeautyCam & Photo & 50M+ & 12.5.35 &  &  &  &  & \\
\rowcolor{gray!15}
4 & Wattpad & Book & 50M+ & 11.2.0 & $\ast$ & $\ast$ & $\ast$ & $\ast$ & $\ast$\\
5 & Strava & Health & 50M+ & 4.2.5 &  &  &  &  & \\
\rowcolor{gray!15}
6 & Klook & Travel & 10M+ & 7.20.2 &  &  &  &  & \\
7 & HelloTalk & Edu & 10M+ & 6.0.6 &  &  &  &  & \\
\rowcolor{gray!15}
8 & Xbrowser & Commun & 10M+ & 5.2.0 &  &  &  &  & \\
9 & MMWbO & Health & 10M+ & 25.2.0 & $\ast$ & $\ast$ &  &  & $\ast$\\
\rowcolor{gray!15}
10 & SportsTrack & Health & 10M+ & 5.0.1 &  &  &  &  & \\
11 & Deliveroo & Food & 10M+ & 3.215 &  &  &  &  & \\
\rowcolor{gray!15}
12 & KiKUU & Shop & 10M+ & 30.1.6 &  & $\ast$ &  &  & \\
13 & Wink & Video & 10M+ & 2.7.5 & $\ast$ &  &  & $\ast$ & $\ast$\\
\rowcolor{gray!15}
14 & StepsApp & Health & 10M+ & 6.0.12 &  &  &  &  & \\
15 & eSound & Music & 10M+ & 6.11 &  &  &  &  & \\
\rowcolor{gray!15}
16 & HabitNow & Product & 5M+ & 2.2.3d &  &  &  &  & \\
17 & Sectograph & Product & 5M+ & 2.3.1 &  &  &  &  & \\
\rowcolor{gray!15}
18 & Comera & Commun & 5M+ & 5.0.23 &  &  &  &  & $\ast$ \\
19 & NetSpeed & Tool & 5M+ & 1.10.0 &  &  &  &  & $\ast$ \\
\rowcolor{gray!15}
20 & Speaky & Edu & 5M+ & 3.3.5 &  &  &  &  & \\
21 & FSBrowser & Social & 5M+ & 8.4.38 & $\ast$ &  & $\ast$ &  & \\
\rowcolor{gray!15}
22 & Sketchbook & Art & 5M+ & 6.10.0 &  &  &  &  & \\
23 & FlipaClip & Art & 5M+ & 4.2.5 &  &  &  &  & \\
\rowcolor{gray!15}
24 & ClickUp & Product & 1M+ & 5.4.6 &  &  &  & $\ast$ & \\
25 & Tawasal & Commun & 1M+ & 5.4.0 &  & $\ast$ &  & $\ast$ & $\ast$\\
\rowcolor{gray!15}
26 & VERO & Social & 1M+ & 2.3.3 &  &  &  &  & \\
27 & Hilokal & Edu & 1M+ & 12.15.3 &  &  &  &  & \\
\rowcolor{gray!15}
28 & 9Weather & Weather & 1M+ & 1.134 &  &  &  & $\ast$ & $\ast$ \\
29 & BEA & Finance & 1M+ & 10.0.1 &  &  &  &  & \\
\rowcolor{gray!15}
30 & Joytify & Music & 1M+ & 1.2.9 &  &  &  &  & \\
31 & Weezer & Music & 1M+ & 2.4.4 &  &  &  &  & \\
\rowcolor{gray!15}
32 & Vyke & Commun & 1M+ & 1.20.3 &  &  &  &  & \\
33 & DailyExpen & Finance & 1M+ & 249 &  &  &  &  & \\
\rowcolor{gray!15}
34 & Supershift & Product & 1M+ & 25.15 &  &  &  &  & \\
35 & Timeleft & Travel & 500K+ & 3.3.0 &  &  &  & \\
\bottomrule
\end{tabular}
\begin{tablenotes}
\footnotesize
\item \textbf{\textit{Notes:}} ``GP'' is the GPTDroid, ``DA'' is the DroidAgent, ``ATA'' is the AUITestAgent, ``VD'' is the VisionDroid, ``GU'' is the Guardian.
\end{tablenotes}
\vspace{-0.1in}
\end{table}

%% file: sec/discussion.tex
\section{Discussion}
\label{sec_discussion}
\subsection{\textbf{Effectiveness Across App Versions}}
We evaluate whether {\tool} can improve GUI testing across different versions of the same app. In practice, mobile apps evolve incrementally through versioned releases, introducing layout refinements, new features, or behavioral changes. To simulate real-world evolution, we select five historical versions (v2.0 to v5.0) of a real-world finance app, Wallet~\cite{Wallet} (Download number: 10M+), covering both minor revisions (e.g., v2.0 to v2.1) and major upgrades (e.g., v2.4 to v4.0/v5.0). At the start of testing, the memory is empty and progressively accumulates experience from earlier versions.


\input{tab/Discussion-1}

As shown in Table \ref{tab:Discussion-1}, we compare the performance of baselines with and without {\tool}. Without memory, baseline coverage remains relatively low and fluctuates across versions (e.g., VisionDroid ranges from 0.29 to 0.35). With memory support, all baselines with {\tool} show improvements as the versions progress. For instance, Guardian with {\tool} improves activity coverage from 0.30 (v2.0) to 0.64 (v5.0), while bug detection increases from 3 to 5. These results suggest that {\tool} effectively supports version-aware testing by reusing past execution knowledge.
Interestingly, performance gains are evident even when the app undergoes major UI changes, such as between v2.4 and v4.0, indicating that {\tool}'s reflective and strategic memory is able to abstract and transfer testing knowledge across moderate structural shifts. This demonstrates the potential of memory-guided testing to enhance regression analysis and ensure robustness in evolving mobile applications.

\subsection{\textbf{Memory Transfer Across Platforms}}
We further evaluate whether the testing memory constructed from Android apps can benefit GUI testing on a different platform (Web). We select two representative web GUI testing tools, SeeAct~\cite{zheng2024seeact} and GBST~\cite{zimmermann2023gui}, and test them on ten real-world Web Apps. We reuse the memory pool constructed from the 60 Android apps used in RQ1, without incorporating any platform-specific knowledge from the web domain.

We compare each Web testing tool's original version with the integrated {\tool} version, using a 60-minute testing time budget per app. The original SeeAct and GBST detect 3 and 1 bugs, respectively. While they integrated with {\tool} detect 7 and 4 bugs. This demonstrates that the memory captured from Android tasks can effectively support web GUI testing by providing generalized testing strategies.
Despite platform differences in UI structure and interaction modality, {\tool} retrieves reflective and strategic memory that captures high-level exploration goals. This helps guide the GUI testing tools toward semantically meaningful actions and reduces ineffective actions. 

\subsection{\textbf{Insights From Memory Content}}
Beyond quantitative improvements, {\tool} produces reflective and strategic memory entries that offer actionable insights into app behavior. These summaries can guide both automated agents and human testers in identifying bugs and optimizing testing strategies.
For example, in one reflective memory, {\tool} observed that in a Spend Tracker~\cite{SpendTracker} app, accessing the ``Add Transaction'' page without selecting a transaction type (e.g., income or expense) and immediately tapping ``Save'' led to a crash. Based on this insight, the GUI testing tool systematically explored variations of this flow in other similar apps, successfully triggering similar bugs. 
Overall, these insights demonstrate that {\tool}'s memory is not merely a static log but a dynamic experience. In future work, we plan to explore how these memory summaries can assist human testers in prioritizing test scenarios, debugging edge cases, and designing testing strategies.

\subsection{\textbf{Threats of Validity}}
\label{subsec_Validity}
The first threat relates to the potential overlap between the apps in the LLM's training data and those used in experiments. Since LLMs are trained on large-scale datasets, including open-source repositories, some of the apps may have been seen during training, potentially influencing the results. To mitigate this threat, we specifically selected apps that were updated after March 2025 for RQ4. This ensures that the apps are new and unseen by the LLM, providing a more robust evaluation of {\tool}.

The second threat is hallucination in LLMs. We provide output examples in the prompt to mitigate its occurrence and ensure the robustness of our {\tool}. 
Despite these mitigation strategies, hallucinations may still occur. We also design pre-defined recovery rules that automatically correct non-executable actions.

%% file: tab/Discussion-1.tex
\begin{table}[h]
\vspace{-0.05in}
\caption{Performance of different versions of an app.}
\renewcommand\arraystretch{1.1} 
\label{tab:Discussion-1}
\centering
\footnotesize
\begin{tabular}{p{2.4cm}<{\centering} | p{0.6cm}<{\centering} | p{0.6cm}<{\centering} | p{0.6cm}<{\centering} | p{0.6cm}<{\centering} | p{0.6cm}<{\centering} | p{0.5cm}<{\centering}}
\toprule
\multirow{2}*{\textbf{Testing Method}} & \multicolumn{5}{c|}{\textbf{Activity Coverage (different versions)}} & \textbf{All} \cr 
& \textbf{v 2.0} & \textbf{v 2.1} & \textbf{v 2.4} & \textbf{v 4.0} & \textbf{v 5.0} &  \textbf{Bugs} \\ 
\midrule
GPTDroid & 0.23  & 0.25  & 0.25 & 0.26 & 0.23 & 1  \\ 
\rowcolor{gray!15}
\textbf{GP+{\tool}} & \textbf{0.23}  & \textbf{0.33}  & \textbf{0.39} & \textbf{0.41} & \textbf{0.43} & \textbf{2}\\ 
\midrule
DroidAgent & 0.27  & 0.26  & 0.28 & 0.26 & 0.29 & 1  \\ 
\rowcolor{gray!15}
\textbf{DA+{\tool}} & \textbf{0.27}  & \textbf{0.35}   & \textbf{0.41} & \textbf{0.41} & \textbf{0.46} & \textbf{2}  \\
\midrule
AUITestAgent & 0.29  & 0.28  & 0.30 & 0.29 & 0.31 & 1  \\ 
\rowcolor{gray!15}
\textbf{ATA+{\tool}} & \textbf{0.29}  & \textbf{0.37}   & \textbf{0.38} & \textbf{0.44} & \textbf{0.49} & \textbf{3}  \\
\midrule
VisionDroid & 0.33  & 0.32  & 0.35 & 0.29 & 0.31 & 2  \\ 
\rowcolor{gray!15}
\textbf{VD+{\tool}} & \textbf{0.33}  & \textbf{0.45}  & \textbf{0.57} & \textbf{0.59} & \textbf{0.59} & \textbf{4}\\ 
\midrule
Guardian & 0.30  & 0.34  & 0.29 & 0.33 & 0.34 & 3  \\ 
\rowcolor{gray!15}
\textbf{GU+{\tool}} & \textbf{0.30}  & \textbf{0.47}   & \textbf{0.53} & \textbf{0.61} & \textbf{0.64} & \textbf{5}  \\
\bottomrule
\end{tabular}
\vspace{0.05in}
\begin{tablenotes}
\footnotesize
\item \textbf{\textit{Notes:}} ``GP'' is the GPTDroid, ``DA'' is the DroidAgent, ``ATA'' is the AUITestAgent, ``VD'' is the VisionDroid, ``GU'' is the Guardian.
\end{tablenotes}
\vspace{-0.05in}
\end{table}

%% file: sec/related.tex
\section{Related Work}
\label{sec_related}
\subsection{\textbf{Automated GUI Testing}}
To ensure the quality of mobile apps, many researchers study the automatic generation of large-scale test scripts to test apps~\cite{xie2007designing}.
Since Android apps are event-based~\cite{anand2012automated,wu2019analyses,jabbarvand2019search,matinnejad2017automated}, the most common automated testing approaches are model-based automated GUI testing approaches~\cite{mirzaei2016reducing,yang2018static,yang2013grey,zeng2016automated,mao2016sapienz,su2017guided,dong2020time,gu2019practical,wang2020combodroid,chen2025standing,liu2024unblind,liu2024make,liu2024testing,liu2023fill,liu2023ex,huang2023context,liu2022navidroid}, which design corresponding models through the analysis of the apps.
Due to the lack of consideration for the semantic information of the app's GUI pages, the coverage of model-based approaches is still low.
Researchers further proposed human-like testing strategies and designed LLM-based automated GUI testing approaches. GPTDroid~\cite{liu2023chatting} used GPT-3.5 to generate the testing script. AppAgent~\cite{yang2023appagent} enabled the agent to operate apps through a simplified action space. 
DroidAgent~\cite{yoon2023autonomous} generated high-level, realistic tasks for GUI testing based on app-specific functionalities.
AUITestAgent~\cite{hu2024auitestagent} is used for step-oriented testing. AXNav~\cite{taeb2024axnav} generated annotated videos to visually and interactively review accessibility test results.
UXAgent~\cite{lu2025uxagent} achieved LLM-driven automated usability testing by simulating thousands of user interactions. TestAgent~\cite{li2024test} generated and executed test cases through multimodal perception.
However, they focus on improving testing performance within single-session testing tasks, without mechanisms for modeling or reusing prior testing experience. In this paper, we design a memory mechanism that enables LLM-based GUI agents to accumulate and reuse testing knowledge.

\subsection{\textbf{Memory Mechanisms in LLM-based Agents}}
Memory emerges as a critical component in enhancing the reasoning and planning capabilities of LLM-based agents, especially in tasks that require multi-step interactions and long-term context~\cite{wang2023augmenting,lu2023memochat,tack2024online}.
In the domain of dialogue agents, recent work introduces persistent memory modules to track long-term user preferences and conversational history~\cite{shao2023character, li2023chatharuhi,wang2023rolellm,zhou2023characterglm,kaiya2023lyfe}, often leveraging memory or embedding-based retrieval to support context retention. Similarly, embodied agents in robotics and game environments~\cite{wang2024jarvis,yan2023larp,wang2023voyager,zhu2023ghost} employ memory to capture past trajectories and outcomes, enabling agents to refine future decisions based on prior experiences.
Software automation agent demonstrated the use of memory in planning complex action sequences by chaining prompts or retrieving execution traces~\cite{lee2023explore,zheng2023synapse}. These approaches typically rely on prompt chaining or retrieval-augmented generation (RAG) to feed historical context into LLMs. However, they treat memory as a history buffer or document corpus, limiting their ability to reason about higher-level task abstraction or reuse strategies across contexts.
Unlike these works, our approach targets memory construction in the specific setting of LLM-based GUI testing, which requires a structured representation of interaction traces and support for cross-task generalization. 

%% file: sec/conclusion.tex
\section{Conclusion}
\label{sec_conclusion}
This paper proposes {\tool}, a three-layer memory mechanism that augments LLM-based GUI testing with the ability to evolve through repeated interaction. These memory layers are dynamically retrieved and injected into LLM prompts at runtime, enabling the agent to reuse successful behaviors, avoid ineffective actions, and prioritize bug-prone paths. We implement {\tool} as a lightweight plugin, which can be integrated into existing LLM-based GUI testing approaches. Evaluation shows that integrating memory improves test coverage and bug detection across diverse apps and platforms. We contribute a reusable memory pool constructed from 60 Android apps, offering a foundation for future research on memory-augmented testing.

In the future, we will expand this memory repository to cover a broader spectrum of apps, with the long-term goal of building a comprehensive and evolving memory base, which could empower agents to generalize better.



%% file: reference.bib
@article{wu2019analyses,
  title={Analyses for specific defects in android applications: A survey},
  author={Wu, Tianyong and Deng, Xi and Yan, Jun and Zhang, Jian},
  journal={Frontiers of Computer Science},
  pages={1--18},
  year={2019},
  publisher={Springer}
}

@inproceedings{anand2012automated,
  title={Automated concolic testing of smartphone apps},
  author={Anand, Saswat and Naik, Mayur and Harrold, Mary Jean and Yang, Hongseok},
  booktitle={Proceedings of the ACM SIGSOFT 20th International Symposium on the Foundations of Software Engineering},
  pages={1--11},
  year={2012}
}

@inproceedings{mao2016sapienz,
  title={Sapienz: Multi-objective automated testing for Android applications},
  author={Mao, Ke and Harman, Mark and Jia, Yue},
  booktitle={Proceedings of the 25th International Symposium on Software Testing and Analysis},
  pages={94--105},
  year={2016}
}

@inproceedings{zeng2016automated,
  title={Automated test input generation for android: Are we really there yet in an industrial case?},
  author={Zeng, Xia and Li, Dengfeng and Zheng, Wujie and Xia, Fan and Deng, Yuetang and Lam, Wing and Yang, Wei and Xie, Tao},
  booktitle={Proceedings of the 2016 24th ACM SIGSOFT International Symposium on Foundations of Software Engineering},
  pages={987--992},
  year={2016}
}

@inproceedings{mirzaei2016reducing,
  title={Reducing combinatorics in GUI testing of android applications},
  author={Mirzaei, Nariman and Garcia, Joshua and Bagheri, Hamid and Sadeghi, Alireza and Malek, Sam},
  booktitle={2016 IEEE/ACM 38th International Conference on Software Engineering (ICSE)},
  pages={559--570},
  year={2016},
  organization={IEEE}
}

@article{yang2018static,
  title={Static window transition graphs for Android},
  author={Yang, Shengqian and Wu, Haowei and Zhang, Hailong and Wang, Yan and Swaminathan, Chandrasekar and Yan, Dacong and Rountev, Atanas},
  journal={Automated Software Engineering},
  volume={25},
  number={4},
  pages={833--873},
  year={2018},
  publisher={Springer}
}

@inproceedings{yang2013grey,
  title={A grey-box approach for automated GUI-model generation of mobile applications},
  author={Yang, Wei and Prasad, Mukul R and Xie, Tao},
  booktitle={International Conference on Fundamental Approaches to Software Engineering},
  pages={250--265},
  year={2013},
  organization={Springer}
}

@misc{Monkey,
  title={Ui/application exerciser monkey},
  author={Developers, Android},
  year={2012}
}

@inproceedings{li2017droidbot,
  title={Droidbot: a lightweight ui-guided test input generator for android},
  author={Li, Yuanchun and Yang, Ziyue and Guo, Yao and Chen, Xiangqun},
  booktitle={2017 IEEE/ACM 39th International Conference on Software Engineering Companion (ICSE-C)},
  pages={23--26},
  year={2017},
  organization={IEEE}
}

@inproceedings{li2019humanoid,
  title={Humanoid: a deep learning-based approach to automated black-box Android app testing},
  author={Li, Yuanchun and Yang, Ziyue and Guo, Yao and Chen, Xiangqun},
  booktitle={2019 34th IEEE/ACM International Conference on Automated Software Engineering (ASE)},
  pages={1070--1073},
  year={2019},
  organization={IEEE}
}

@article{chen2025standing,
  title={Standing on the Shoulders of Giants: Bug-Aware Automated GUI Testing via Retrieval Augmentation},
  author={Chen, Mengzhuo and Liu, Zhe and Chen, Chunyang and Wang, Junjie and Wu, Boyu and Hu, Jun and Wang, Qing},
  journal={Proceedings of the ACM on Software Engineering},
  volume={2},
  number={FSE},
  pages={825--846},
  year={2025},
  publisher={ACM New York, NY, USA}
}

@inproceedings{liu2024unblind,
  title={Unblind text inputs: predicting hint-text of text input in mobile apps via LLM},
  author={Liu, Zhe and Chen, Chunyang and Wang, Junjie and Chen, Mengzhuo and Wu, Boyu and Huang, Yuekai and Hu, Jun and Wang, Qing},
  booktitle={Proceedings of the 2024 CHI Conference on Human Factors in Computing Systems},
  pages={1--20},
  year={2024}
}

@inproceedings{liu2024make,
  title={Make llm a testing expert: Bringing human-like interaction to mobile gui testing via functionality-aware decisions},
  author={Liu, Zhe and Chen, Chunyang and Wang, Junjie and Chen, Mengzhuo and Wu, Boyu and Che, Xing and Wang, Dandan and Wang, Qing},
  booktitle={Proceedings of the IEEE/ACM 46th International Conference on Software Engineering},
  pages={1--13},
  year={2024}
}

@inproceedings{liu2024testing,
  title={Testing the limits: Unusual text inputs generation for mobile app crash detection with large language model},
  author={Liu, Zhe and Chen, Chunyang and Wang, Junjie and Chen, Mengzhuo and Wu, Boyu and Tian, Zhilin and Huang, Yuekai and Hu, Jun and Wang, Qing},
  booktitle={Proceedings of the IEEE/ACM 46th International conference on software engineering},
  pages={1--12},
  year={2024}
}

@inproceedings{liu2023fill,
  title={Fill in the blank: Context-aware automated text input generation for mobile gui testing},
  author={Liu, Zhe and Chen, Chunyang and Wang, Junjie and Che, Xing and Huang, Yuekai and Hu, Jun and Wang, Qing},
  booktitle={2023 IEEE/ACM 45th International Conference on Software Engineering (ICSE)},
  pages={1355--1367},
  year={2023},
  organization={IEEE}
}

@inproceedings{liu2023ex,
  title={Ex pede herculem: Augmenting activity transition graph for apps via graph convolution network},
  author={Liu, Zhe and Chen, Chunyang and Wang, Junjie and Su, Yuhui and Huang, Yuekai and Hu, Jun and Wang, Qing},
  booktitle={2023 IEEE/ACM 45th International Conference on Software Engineering (ICSE)},
  pages={1983--1995},
  year={2023},
  organization={IEEE}
}

@inproceedings{huang2023context,
  title={Context-aware bug reproduction for mobile apps},
  author={Huang, Yuchao and Wang, Junjie and Liu, Zhe and Wang, Song and Chen, Chunyang and Li, Mingyang and Wang, Qing},
  booktitle={2023 IEEE/ACM 45th International Conference on Software Engineering (ICSE)},
  pages={2336--2348},
  year={2023},
  organization={IEEE}
}

@inproceedings{liu2022navidroid,
  title={NaviDroid: a tool for guiding manual Android testing via hint moves},
  author={Liu, Zhe and Chen, Chunyang and Wang, Junjie and Su, Yuhui and Wang, Qing},
  booktitle={Proceedings of the ACM/IEEE 44th international conference on software engineering: companion proceedings},
  pages={154--158},
  year={2022}
}

@inproceedings{liu2022guided,
  title={Guided bug crush: Assist manual gui testing of android apps via hint moves},
  author={Liu, Zhe and Chen, Chunyang and Wang, Junjie and Huang, Yuekai and Hu, Jun and Wang, Qing},
  booktitle={Proceedings of the 2022 CHI conference on human factors in computing systems},
  pages={1--14},
  year={2022}
}

@inproceedings{su2017guided,
	title={Guided, stochastic model-based GUI testing of Android apps},
	author={Su, Ting and Meng, Guozhu and Chen, Yuting and Wu, Ke and Yang, Weiming and Yao, Yao and Pu, Geguang and Liu, Yang and Su, Zhendong},
	booktitle={Proceedings of the 2017 11th Joint Meeting on Foundations of Software Engineering},
	pages={245--256},
	year={2017}
}

@Misc{Activity,
    title = {Android documentation: Activity.)},
	howpublished = {\url{https://developer.android.com/reference/android/app/Activity}},	
	year={2020},	
}

@inproceedings{gu2019practical,
  title={Practical GUI testing of Android applications via model abstraction and refinement},
  author={Gu, Tianxiao and Sun, Chengnian and Ma, Xiaoxing and Cao, Chun and Xu, Chang and Yao, Yuan and Zhang, Qirun and Lu, Jian and Su, Zhendong},
  booktitle={2019 IEEE/ACM 41st International Conference on Software Engineering (ICSE)},
  pages={269--280},
  year={2019},
  organization={IEEE}
}

@inproceedings{pan2020reinforcement,
  title={Reinforcement learning based curiosity-driven testing of Android applications},
  author={Pan, Minxue and Huang, An and Wang, Guoxin and Zhang, Tian and Li, Xuandong},
  booktitle={Proceedings of the 29th ACM SIGSOFT International Symposium on Software Testing and Analysis},
  pages={153--164},
  year={2020}
}

@Misc{Android,
    title = {Android},
	howpublished = {\url{https://developer.android.google/topic/}},	
	year={2022},	
}

@inproceedings{he2020textexerciser,
  title={TextExerciser: feedback-driven text input exercising for android applications},
  author={He, Yuyu and Zhang, Lei and Yang, Zhemin and Cao, Yinzhi and Lian, Keke and Li, Shuai and Yang, Wei and Zhang, Zhibo and Yang, Min and Zhang, Yuan and others},
  booktitle={2020 IEEE Symposium on Security and Privacy (SP)},
  pages={1071--1087},
  year={2020},
  organization={IEEE}
}

@article{arnatovich2018mobolic,
  title={Mobolic: An automated approach to exercising mobile application GUIs using symbiosis of online testing technique and customated input generation},
  author={Arnatovich, Yauhen Leanidavich and Wang, Lipo and Ngo, Ngoc Minh and Soh, Charlie},
  journal={Software: Practice and Experience},
  volume={48},
  number={5},
  pages={1107--1142},
  year={2018},
  publisher={Wiley Online Library}
}

@inproceedings{liu2017automatic,
  title={Automatic text input generation for mobile testing},
  author={Liu, Peng and Zhang, Xiangyu and Pistoia, Marco and Zheng, Yunhui and Marques, Manoel and Zeng, Lingfei},
  booktitle={2017 IEEE/ACM 39th International Conference on Software Engineering (ICSE)},
  pages={643--653},
  year={2017},
  organization={IEEE}
}

@inproceedings{wang2020combodroid,
  title={Combodroid: generating high-quality test inputs for android apps via use case combinations},
  author={Wang, Jue and Jiang, Yanyan and Xu, Chang and Cao, Chun and Ma, Xiaoxing and Lu, Jian},
  booktitle={ICSE},
  pages={469--480},
  year={2020}
}

@inproceedings{dong2020time,
  title={Time-travel testing of android apps},
  author={Dong, Zhen and B{\"o}hme, Marcel and Cojocaru, Lucia and Roychoudhury, Abhik},
  booktitle={2020 IEEE/ACM 42nd International Conference on Software Engineering (ICSE)},
  pages={481--492},
  year={2020},
  organization={IEEE}
}

@inproceedings{su2021benchmarking,
  title={Benchmarking automated GUI testing for Android against real-world bugs},
  author={Su, Ting and Wang, Jue and Su, Zhendong},
  booktitle={Proceedings of the 29th ACM Joint Meeting on European Software Engineering Conference and Symposium on the Foundations of Software Engineering},
  pages={119--130},
  year={2021}
}

@inproceedings{wang2021vet,
  title={Vet: identifying and avoiding UI exploration tarpits},
  author={Wang, Wenyu and Yang, Wei and Xu, Tianyin and Xie, Tao},
  booktitle={FSE},
  pages={83--94},
  year={2021}
}

@Misc{uiautomator,
  title={Python wrapper of Android uiautomator test tool.},
  author={UIAutomator},
  howpublished = {\url{https://github.
com/xiaocong/uiautomator}},	
  year={2021}
}

@article{liu2023chatting,
  title={Chatting with gpt-3 for zero-shot human-like mobile automated gui testing},
  author={Liu, Zhe and Chen, Chunyang and Wang, Junjie and Chen, Mengzhuo and Wu, Boyu and Che, Xing and Wang, Dandan and Wang, Qing},
  journal={arXiv preprint arXiv:2305.09434},
  year={2023}
}

@article{xie2007designing,
  title={Designing and comparing automated test oracles for GUI-based software applications},
  author={Xie, Qing and Memon, Atif M},
  journal={ACM Transactions on Software Engineering and Methodology (TOSEM)},
  volume={16},
  number={1},
  pages={4--es},
  year={2007},
  publisher={ACM New York, NY, USA}
}

@Misc{github,
        title={GitHub},
	howpublished = {\url{https://github.com/}},	
	year={2024},	
}

@article{yang2023appagent,
  title={Appagent: Multimodal agents as smartphone users},
  author={Yang, Zhao and Liu, Jiaxuan and Han, Yucheng and Chen, Xin and Huang, Zebiao and Fu, Bin and Yu, Gang},
  journal={arXiv preprint arXiv:2312.13771},
  year={2023}
}

@inproceedings{wen2024autodroid,
  title={AutoDroid: LLM-powered Task Automation in Android},
  author={Wen, Hao and Li, Yuanchun and Liu, Guohong and Zhao, Shanhui and Yu, Tao and Li, Toby Jia-Jun and Jiang, Shiqi and Liu, Yunhao and Zhang, Yaqin and Liu, Yunxin},
  booktitle={Proceedings of the 30th Annual International Conference on Mobile Computing and Networking},
  pages={543--557},
  year={2024}
}

@inproceedings{matinnejad2017automated,
  title={Automated testing of hybrid Simulink/Stateflow controllers: industrial case studies},
  author={Matinnejad, Reza and Nejati, Shiva and Briand, Lionel C},
  booktitle={Proceedings of the 2017 11th Joint Meeting on Foundations of Software Engineering},
  pages={938--943},
  year={2017}
}

@inproceedings{jabbarvand2019search,
  title={Search-based energy testing of android},
  author={Jabbarvand, Reyhaneh and Lin, Jun-Wei and Malek, Sam},
  booktitle={ICSE},
  pages={1119--1130},
  year={2019},
  organization={IEEE}
}

@article{yasin2021droidbotx,
  title={Droidbotx: Test case generation tool for android applications using Q-learning},
  author={Yasin, Husam N and Hamid, Siti Hafizah Ab and Raja Yusof, Raja Jamilah},
  journal={Symmetry},
  year={2021}
}

@inproceedings{su2024enhancing,
  title={Enhancing exploratory testing by large language model and knowledge graph},
  author={Su, Yanqi and Liao, Dianshu and Xing, Zhenchang and Huang, Qing and Xie, Mulong and Lu, Qinghua and Xu, Xiwei},
  booktitle={Proceedings of the IEEE/ACM 46th International Conference on Software Engineering},
  pages={1--12},
  year={2024}
}

@inproceedings{su2022constructing,
  title={Constructing a system knowledge graph of user tasks and failures from bug reports to support soap opera testing},
  author={Su, Yanqi and Han, Zheming and Xing, Zhenchang and Xia, Xin and Xu, Xiwei and Zhu, Liming and Lu, Qinghua},
  booktitle={Proceedings of the 37th IEEE/ACM International Conference on Automated Software Engineering},
  pages={1--13},
  year={2022}
}

@inproceedings{wang2024knowledge,
  title={Knowledge Graph Driven Inference Testing for Question Answering Software},
  author={Wang, Jun and Li, Yanhui and Chen, Zhifei and Chen, Lin and Zhang, Xiaofang and Zhou, Yuming},
  booktitle={Proceedings of the IEEE/ACM 46th International Conference on Software Engineering},
  pages={1--13},
  year={2024}
}

@article{wang2024mobile,
  title={Mobile-agent: Autonomous multi-modal mobile device agent with visual perception},
  author={Wang, Junyang and Xu, Haiyang and Ye, Jiabo and Yan, Ming and Shen, Weizhou and Zhang, Ji and Huang, Fei and Sang, Jitao},
  journal={arXiv preprint arXiv:2401.16158},
  year={2024}
}

@article{yoon2023autonomous,
  title={Autonomous large language model agents enabling intent-driven mobile gui testing},
  author={Yoon, Juyeon and Feldt, Robert and Yoo, Shin},
  journal={arXiv preprint arXiv:2311.08649},
  year={2023}
}

@article{hu2024auitestagent,
  title={Auitestagent: Automatic requirements oriented gui function testing},
  author={Hu, Yongxiang and Wang, Xuan and Wang, Yingchuan and Zhang, Yu and Guo, Shiyu and Chen, Chaoyi and Wang, Xin and Zhou, Yangfan},
  journal={arXiv preprint arXiv:2407.09018},
  year={2024}
}

@article{liu2024vision,
  title={Vision-driven automated mobile gui testing via multimodal large language model},
  author={Liu, Zhe and Li, Cheng and Chen, Chunyang and Wang, Junjie and Wu, Boyu and Wang, Yawen and Hu, Jun and Wang, Qing},
  journal={arXiv preprint arXiv:2407.03037},
  year={2024}
}

@article{zhang2023appagent,
  title={Appagent: Multimodal agents as smartphone users},
  author={Zhang, Chi and Yang, Zhao and Liu, Jiaxuan and Han, Yucheng and Chen, Xin and Huang, Zebiao and Fu, Bin and Yu, Gang},
  journal={arXiv preprint arXiv:2312.13771},
  year={2023}
}

@article{li2024appagent,
  title={Appagent v2: Advanced agent for flexible mobile interactions},
  author={Li, Yanda and Zhang, Chi and Yang, Wanqi and Fu, Bin and Cheng, Pei and Chen, Xin and Chen, Ling and Wei, Yunchao},
  journal={arXiv preprint arXiv:2408.11824},
  year={2024}
}

@article{wen2023droidbot,
  title={Droidbot-gpt: Gpt-powered ui automation for android},
  author={Wen, Hao and Wang, Hongming and Liu, Jiaxuan and Li, Yuanchun},
  journal={arXiv preprint arXiv:2304.07061},
  year={2023}
}

@inproceedings{ran2024guardian,
  title={Guardian: A Runtime Framework for LLM-Based UI Exploration},
  author={Ran, Dezhi and Wang, Hao and Song, Zihe and Wu, Mengzhou and Cao, Yuan and Zhang, Ying and Yang, Wei and Xie, Tao},
  booktitle={Proceedings of the 33rd ACM SIGSOFT International Symposium on Software Testing and Analysis},
  pages={958--970},
  year={2024}
}

@article{itkonen2015test,
  title={Test better by exploring: Harnessing human skills and knowledge},
  author={Itkonen, Juha and M{\"a}ntyl{\"a}, Mika V and Lassenius, Casper},
  journal={IEEE Software},
  volume={33},
  number={4},
  pages={90--96},
  year={2015},
  publisher={IEEE}
}

@article{shao2023character,
  title={Character-llm: A trainable agent for role-playing},
  author={Shao, Yunfan and Li, Linyang and Dai, Junqi and Qiu, Xipeng},
  journal={arXiv preprint arXiv:2310.10158},
  year={2023}
}

@article{li2023metaagents,
  title={Metaagents: Simulating interactions of human behaviors for llm-based task-oriented coordination via collaborative generative agents},
  author={Li, Yuan and Zhang, Yixuan and Sun, Lichao},
  journal={arXiv preprint arXiv:2310.06500},
  year={2023}
}

@article{craik1992human,
  title={Human memory.},
  author={Craik, Fergus IM and Jennings, Janine M},
  year={1992},
  publisher={Lawrence Erlbaum Associates, Inc}
}

@book{baddeley1997human,
  title={Human memory: Theory and practice},
  author={Baddeley, Alan D},
  year={1997},
  publisher={psychology press}
}

@inproceedings{Radford2021LearningTV,
  title={Learning Transferable Visual Models From Natural Language Supervision},
  author={Alec Radford and Jong Wook Kim and Chris Hallacy and A. Ramesh and Gabriel Goh and Sandhini Agarwal and Girish Sastry and Amanda Askell and Pamela Mishkin and Jack Clark and Gretchen Krueger and Ilya Sutskever},
  booktitle={ICML},
  year={2021}
}

@article{lu2023memochat,
  title={Memochat: Tuning llms to use memos for consistent long-range open-domain conversation},
  author={Lu, Junru and An, Siyu and Lin, Mingbao and Pergola, Gabriele and He, Yulan and Yin, Di and Sun, Xing and Wu, Yunsheng},
  journal={arXiv preprint arXiv:2308.08239},
  year={2023}
}

@article{packer2023memgpt,
  title={MemGPT: Towards LLMs as Operating Systems.},
  author={Packer, Charles and Fang, Vivian and Patil, Shishir\_G and Lin, Kevin and Wooders, Sarah and Gonzalez, Joseph\_E},
  year={2023},
  publisher={arXiv}
}

@article{sumers2023cognitive,
  title={Cognitive architectures for language agents},
  author={Sumers, Theodore and Yao, Shunyu and Narasimhan, Karthik and Griffiths, Thomas},
  journal={Transactions on Machine Learning Research},
  year={2023}
}

@inproceedings{tsai2024rtlfixer,
  title={RTLFixer: Automatically fixing RTL syntax errors with large language model},
  author={Tsai, YunDa and Liu, Mingjie and Ren, Haoxing},
  booktitle={Proceedings of the 61st ACM/IEEE Design Automation Conference},
  pages={1--6},
  year={2024}
}

@article{chen2023gamegpt,
  title={Gamegpt: Multi-agent collaborative framework for game development},
  author={Chen, Dake and Wang, Hanbin and Huo, Yunhao and Li, Yuzhao and Zhang, Haoyang},
  journal={arXiv preprint arXiv:2310.08067},
  year={2023}
}

@article{xie2024embodied,
  title={Embodied-RAG: General non-parametric embodied memory for retrieval and generation},
  author={Xie, Quanting and Min, So Yeon and Ji, Pengliang and Yang, Yue and Zhang, Tianyi and Xu, Kedi and Bajaj, Aarav and Salakhutdinov, Ruslan and Johnson-Roberson, Matthew and Bisk, Yonatan},
  journal={arXiv preprint arXiv:2409.18313},
  year={2024}
}

@article{mei2024continuously,
  title={Continuously learning, adapting, and improving: A dual-process approach to autonomous driving},
  author={Mei, Jianbiao and Ma, Yukai and Yang, Xuemeng and Wen, Licheng and Cai, Xinyu and Li, Xin and Fu, Daocheng and Zhang, Bo and Cai, Pinlong and Dou, Min and others},
  journal={arXiv preprint arXiv:2405.15324},
  year={2024}
}

@misc{Wallet,
  title={Wallet},
   howpublished = {\url{https://play.google.com/store/apps/details?id=com.droid4you.application.wallet}},
   year = {2025}
}

@inproceedings{zimmermann2023gui,
  title={Gui-based software testing: An automated approach using gpt-4 and selenium webdriver},
  author={Zimmermann, Daniel and Koziolek, Anne},
  booktitle={2023 38th IEEE/ACM International Conference on Automated Software Engineering Workshops (ASEW)},
  pages={171--174},
  year={2023},
  organization={IEEE}
}

@inproceedings{zheng2024seeact,
  title={GPT-4V(ision) is a Generalist Web Agent, if Grounded},
  author={Boyuan Zheng and Boyu Gou and Jihyung Kil and Huan Sun and Yu Su},
  booktitle={Forty-first International Conference on Machine Learning},
  year={2024},
  url={https://openreview.net/forum?id=piecKJ2DlB},
}

@misc{SpendTracker,
  title={Spend Tracker},
   howpublished = {\url{https://play.google.com/store/apps/details?id=com.thebudgetingapp.thebudgetingapp&hl=en}},
   year = {2025}
}

@inproceedings{taeb2024axnav,
  title={Axnav: Replaying accessibility tests from natural language},
  author={Taeb, Maryam and Swearngin, Amanda and Schoop, Eldon and Cheng, Ruijia and Jiang, Yue and Nichols, Jeffrey},
  booktitle={Proceedings of the 2024 CHI Conference on Human Factors in Computing Systems},
  pages={1--16},
  year={2024}
}

@article{lu2025uxagent,
  title={Uxagent: An llm agent-based usability testing framework for web design},
  author={Lu, Yuxuan and Yao, Bingsheng and Gu, Hansu and Huang, Jing and Wang, Jessie and Li, Yang and Gesi, Jiri and He, Qi and Li, Toby Jia-Jun and Wang, Dakuo},
  journal={arXiv preprint arXiv:2502.12561},
  year={2025}
}

@inproceedings{li2024test,
  title={Test-Agent: A Multimodal App Automation Testing Framework Based on the Large Language Model},
  author={Li, Youwei and Li, Yangyang and Yang, Yangzhao},
  booktitle={2024 IEEE 4th International Conference on Digital Twins and Parallel Intelligence (DTPI)},
  pages={609--614},
  year={2024},
  organization={IEEE}
}

@article{li2023chatharuhi,
  title={Chatharuhi: Reviving anime character in reality via large language model},
  author={Li, Cheng and Leng, Ziang and Yan, Chenxi and Shen, Junyi and Wang, Hao and Mi, Weishi and Fei, Yaying and Feng, Xiaoyang and Yan, Song and Wang, HaoSheng and others},
  journal={arXiv preprint arXiv:2308.09597},
  year={2023}
}

@article{wang2023rolellm,
  title={Rolellm: Benchmarking, eliciting, and enhancing role-playing abilities of large language models},
  author={Wang, Zekun Moore and Peng, Zhongyuan and Que, Haoran and Liu, Jiaheng and Zhou, Wangchunshu and Wu, Yuhan and Guo, Hongcheng and Gan, Ruitong and Ni, Zehao and Yang, Jian and others},
  journal={arXiv preprint arXiv:2310.00746},
  year={2023}
}

@article{zhou2023characterglm,
  title={Characterglm: Customizing chinese conversational ai characters with large language models},
  author={Zhou, Jinfeng and Chen, Zhuang and Wan, Dazhen and Wen, Bosi and Song, Yi and Yu, Jifan and Huang, Yongkang and Peng, Libiao and Yang, Jiaming and Xiao, Xiyao and others},
  journal={arXiv preprint arXiv:2311.16832},
  year={2023}
}

@article{kaiya2023lyfe,
  title={Lyfe agents: Generative agents for low-cost real-time social interactions},
  author={Kaiya, Zhao and Naim, Michelangelo and Kondic, Jovana and Cortes, Manuel and Ge, Jiaxin and Luo, Shuying and Yang, Guangyu Robert and Ahn, Andrew},
  journal={arXiv preprint arXiv:2310.02172},
  year={2023}
}

@article{wang2024jarvis,
  title={Jarvis-1: Open-world multi-task agents with memory-augmented multimodal language models},
  author={Wang, Zihao and Cai, Shaofei and Liu, Anji and Jin, Yonggang and Hou, Jinbing and Zhang, Bowei and Lin, Haowei and He, Zhaofeng and Zheng, Zilong and Yang, Yaodong and others},
  journal={IEEE Transactions on Pattern Analysis and Machine Intelligence},
  year={2024},
  publisher={IEEE}
}

@article{yan2023larp,
  title={Larp: Language-agent role play for open-world games},
  author={Yan, Ming and Li, Ruihao and Zhang, Hao and Wang, Hao and Yang, Zhilan and Yan, Ji},
  journal={arXiv preprint arXiv:2312.17653},
  year={2023}
}

@article{wang2023voyager,
  title={Voyager: An open-ended embodied agent with large language models},
  author={Wang, Guanzhi and Xie, Yuqi and Jiang, Yunfan and Mandlekar, Ajay and Xiao, Chaowei and Zhu, Yuke and Fan, Linxi and Anandkumar, Anima},
  journal={arXiv preprint arXiv:2305.16291},
  year={2023}
}

@article{zhu2023ghost,
  title={Ghost in the minecraft: Generally capable agents for open-world environments via large language models with text-based knowledge and memory},
  author={Zhu, Xizhou and Chen, Yuntao and Tian, Hao and Tao, Chenxin and Su, Weijie and Yang, Chenyu and Huang, Gao and Li, Bin and Lu, Lewei and Wang, Xiaogang and others},
  journal={arXiv preprint arXiv:2305.17144},
  year={2023}
}

@article{lee2023explore,
  title={Explore, select, derive, and recall: Augmenting llm with human-like memory for mobile task automation},
  author={Lee, Sunjae and Choi, Junyoung and Lee, Jungjae and Wasi, Munim Hasan and Choi, Hojun and Ko, Steven Y and Oh, Sangeun and Shin, Insik},
  journal={arXiv preprint arXiv:2312.03003},
  year={2023}
}

@article{wang2023augmenting,
  title={Augmenting language models with long-term memory},
  author={Wang, Weizhi and Dong, Li and Cheng, Hao and Liu, Xiaodong and Yan, Xifeng and Gao, Jianfeng and Wei, Furu},
  journal={Advances in Neural Information Processing Systems},
  volume={36},
  pages={74530--74543},
  year={2023}
}

@article{tack2024online,
  title={Online adaptation of language models with a memory of amortized contexts},
  author={Tack, Jihoon and Kim, Jaehyung and Mitchell, Eric and Shin, Jinwoo and Teh, Yee Whye and Schwarz, Jonathan Richard},
  journal={arXiv preprint arXiv:2403.04317},
  year={2024}
}

@article{zheng2023synapse,
  title={Synapse: Trajectory-as-exemplar prompting with memory for computer control},
  author={Zheng, Longtao and Wang, Rundong and Wang, Xinrun and An, Bo},
  journal={arXiv preprint arXiv:2306.07863},
  year={2023}
}
